\documentclass[a4paper,10pt,twocolumn,nobalancelastpage,aps,pra,superscriptaddress,nofootinbib,longbibliography
]{revtex4-1}

\usepackage[english]{babel}
\usepackage[utf8]{inputenc}
\usepackage[T1]{fontenc}
\usepackage[kerning=true,protrusion=true,expansion=true]{microtype}

\usepackage{amssymb,amsmath,bm}

\usepackage{graphicx}
\usepackage{color,framed,mdframed}
\usepackage[dvipsnames]{xcolor}
\graphicspath{{./fig/}{./figs/}{./tikz/}}

\usepackage[pdftex,
	colorlinks,
	allcolors=blue,
	pdfauthor={A. Rakhubovsky},
	pdftitle={},
	pdfsubject={},
	pdfkeywords={},
	pdfproducer={Latex with hyperref},
	pdfcreator={pdflatex}]{hyperref}

\usepackage[capitalize]{cleveref}

\newcommand*{\s}[1]{\ensuremath{_\text{#1}}}
\newcommand*{\up}[1]{\ensuremath{^\text{#1}}}
\newcommand*{\da}{\dagger}
\newcommand*{\dd}[1]{\ensuremath{\mathrm{d}#1\:}}
\newcommand*{\comm}[2]{\ensuremath{ \left[ #1 , #2 \right] }}
\newcommand*{\acomm}[2]{\ensuremath{ \left\{ #1 , #2 \right\} }}
\newcommand*{\avg}[1]{\left\langle #1 \right\rangle}
\newcommand*{\smtr}[2]{\ensuremath{\avg{ #1 \circ #2 }}}
\newcommand*{\cG}{\ensuremath{\mathfrak{G}}}
\newcommand*{\Mode}[1]{\ensuremath{\mathcal{#1}}}
\renewcommand*{\vec}{\bm}
\newcommand*{\mat}[1]{\ensuremath{\mathbb{#1}}}
\DeclareMathOperator*{\Tr}{\operatorname{Tr}}
\DeclareMathOperator*{\diag}{\operatorname{diag}}

\begin{document}

\title{Nonclassical states of levitated macroscopic objects beyond the ground state}
\author{Andrey A. \surname{Rakhubovsky}}
\email{andrey.rakhubovsky@gmail.com}
\author{Darren W. \surname{Moore}}
\author{Radim \surname{Filip}}
\affiliation{Department of Optics, Palack{\'y} University, 17. Listopadu 12, 771 46 Olomouc, Czech Republic }

\begin{abstract}
	The preparation of nonclassical states of mechanical motion conclusively proves that control over such motion has reached the quantum level.
	We investigate ways to achieve nonclassical states of macroscopic mechanical oscillators, particularly levitated nanoparticles.
	We analyze the possibility of the conditional squeezing of the levitated particle induced by the homodyne detection of light in a pulsed optomechanical setup within the resolved sideband regime.
	We focus on the regimes that are experimentally relevant for the levitated systems where the ground-state cooling is not achievable and the optomechanical coupling is comparable with the cavity linewidth.
	The analysis is thereby performed beyond the adiabatic regime routinely used for the bulk optomechanical pulsed systems.
	The results show that the quantum state of a levitated particle could be squeezed below the ground state variance within a wide range of temperatures.
	This opens a path to test for the first time nonclassical control of levitating nanoparticles beyond the ground state.
\end{abstract}

\date{\today}
\maketitle


\section{Introduction} 
\label{sec:introduction}

Optomechanics~\cite{ meystre_short_2013, *aspelmeyer_cavity_2014, *khalili_quantum_2016} is a field studying systems in which a light or microwave mode is coupled to mechanical motion via radiation pressure.
It has developed at dramatic rates recently, particularly in providing access to strong coupling~\cite{groblacher_observation_2009,*verhagen_quantum-coherent_2012}, optomechanically induced transparency~\cite{weis_optomechanically_2010}, ground state cooling~\cite{chan_laser_2011,*teufel_sideband_2011}, mechanical squeezing~\cite{wollman_quantum_2015,pirkkalainen_squeezing_2015}, nonclassical correlations between photons and phonons~\cite{riedinger_non-classical_2016,*hong_hanbury_2017}, and optomechanical entanglement~\cite{palomaki_entangling_2013,riedinger_remote_2018,ockeloen-korppi_stabilized_2018}, all of which have been demonstrated in the lab.
The remaining step would be to prove that control of the mechanics reaches a level incompatible with any mixture of classical motional states.
To prepare and verify nonclassical aspects, a pulsed scheme~\cite{hofer_quantum_2011,vanner_pulsed_2011} would be advantageous.
It unambiguously separates state preparation and verification~\cite{palomaki_entangling_2013} and keeps the optomechanical system out of the unstable regime inherently peculiar to the optomechanical systems with a single continuous-wave pump~\cite{braginsky_notitle_1964,*braginsky_ponderomotive_1967}.

There is great interest in achieving nonclassical states of mechanics for enhancing metrological performance including force~\cite{rugar_single_2004, gavartin_hybrid_2012}, mass~\cite{seveso_quantum_2017} and displacement sensing~\cite{Abbott:2007kv,*ligo_scientific_collaboration_and_virgo_collaboration_observation_2016}, magnetometry~\cite{forstner_cavity_2012,*yu_optomechanical_2016,*li_quantum_2018}, and biological applications~\cite{arndt_quantum_2009}.
Optomechanical systems find application in the field of quantum information, particularly to transduce~\cite{bagci_optical_2014,*andrews_bidirectional_2014,*andrews_quantum-enabled_2015,*lecocq_mechanically_2016} and route~\cite{peterson_demonstration_2017,*barzanjeh_mechanical_2017,*malz_quantum-limited_2018,*ruesink_optical_2018} quantum signals.
Moreover, a macroscopic object is of great use for testing the fundamental validity of quantum mechanics at larger mass scales~\cite{kaltenbaek_macroscopic_2016,lei_quantum_2016,santos_optomechanical_2017} and probing decoherence models~\cite{romero-isart_quantum_2011,*bera_proposal_2015,*vinante_improved_2017}.

An excellent candidate for many applications happens to be optomechanics with levitated nanoparticles~\cite{chang_cavity_2010,*romero-isart_optically_2011,barker_doppler_2010,kiesel_cavity_2013,yin_optomechanics_2013}.
Levitated particles show great isolation from the environment and therefore, owing to the elimination of the clamping losses, exceptionally high mechanical Q-factors.
The oscillatory motion of the nanoparticle is provided by the optical trap, and can therefore be engineered with great precision.
Moreover, the optical potential for the nanoparticle can be adjusted to be nonlinear~\cite{dholakia_shaping_2011,gieseler_thermal_2013,fonseca_nonlinear_2016,ricci_optically_2017,siler_thermally_2017,siler_diffusing_2018} and variable in time~\cite{konopik_nonequilibrium_2018}.
Simultaneously, it can be controlled in the sideband resolved regime~\cite{kiesel_cavity_2013,millen_cavity_2015} with full control of the linearized interaction between light and nanoparticle motion.
At the same time, cooling the nanoparticles to the ground state is still challenging~\cite{gieseler_subkelvin_2012,asenbaum_cavity_2013,millen_cavity_2015,jain_direct_2016}.
Until now, only classical squashing has been demonstrated with levitating nanospheres~\cite{rashid_experimental_2016}.

In this paper we investigate the possibility for the linearized optomechanical interaction to create quantum correlations between a levitated particle and a control field.
We show that these correlations are sufficient to steer the levitated particle into a conditionally squeezed state upon detection of the optical mode.
To reach this, we use an amplifier interaction in the resolved sideband regime (at the blue sideband), however, we have to go beyond the adiabatic elimination of the intracavity field as the optomechanical coupling in the levitated systems in non-negligible with respect to the cavity decay.
Therefore, we have to optimize the temporal modes of light beyond the frequently used exponential time profiles~\cite{hofer_quantum_2011,palomaki_entangling_2013,andrews_quantum-enabled_2015}.
After this optimization, the measurement-induced preparation of mechanical squeezed states is tolerant to the occupations of the mechanical environment reaching $\sim 10^7$ phonons, what corresponds to the average occupation of a~$100$~kHz oscillator at~$50$~K.

\section{Pulsed Sideband-resolved Optomechanics} 
\label{sec:the_system_and_the_hamiltonian}

\begin{figure}[ht]
	\centering
	\includegraphics[width=0.8\linewidth]{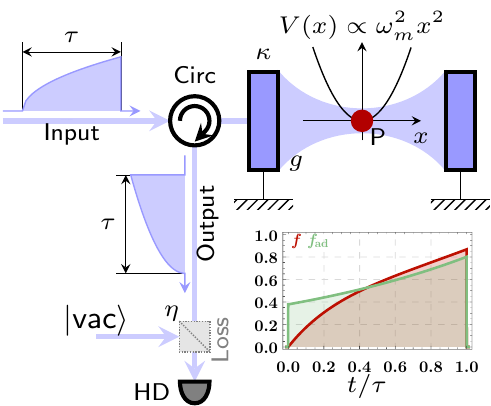}
	\caption{A sketch of the measurement-induced mechanical squeezing of levitated nanoparticle.
		The nanoparticle P is trapped in a harmonic potential by an auxiliary trapping beam (not shown).
		The motion of the particle is coupled to a mode of the cavity at rate $g$.
		The cavity is illuminated by a pulse (Input).
		In the presence of a strong blue-detuned pump the motion of the particle becomes correlated with a particular temporal mode of the leaking field (Output) directed via the circulator (Circ) to the homodyne detector (HD).
		Homodyne detection can project the mechanical system onto a displaced thermal squeezed state.
		Inset: example of the different mode shapes.
		Red (darker) line corresponds to the optimal mode defined by~\eqref{eq:def_out_mode}, green (lighter) to the approximate exponential modes~\eqref{eq:exponential_pulses}.
	}
	\label{fig:fig0:scheme}
\end{figure}

In this section, following the usual textbook approach~\cite{bowen_quantum_2015}, we derive the equations of motion for the effective two-mode squeezing interaction in an optomechanical system.

We consider a standard optomechanical system~\cite{aspelmeyer_cavity_2014} which has an optical mode coupled to the centre of mass motion of a levitated particle via radiation pressure (see Fig.~\ref{fig:fig0:scheme}).
The mechanical motion modulates the cavity frequency, which gives rise to the standard optomechanical Hamiltonian (we use $\hbar = 1$)~\cite{law_interaction_1995,chang_cavity_2010,*romero-isart_optically_2011,monteiro_dynamics_2013}
\begin{equation}
	H = [ \omega\s{cav} - g_0 (a_m^\da + a_m )] a_c^\da a_c + \omega_m a_m^\da a_m + H\s{pump},
\end{equation}
where $a_c(a_m)$ is the annihilation operator of the cavity (mechanical) mode with the eigenfrequency $\omega\s{cav}(\omega_m)$, and $g_0$ is the single-photon optomechanical coupling rate  approximately showing the cavity frequency modulation per mechanical quanta.
We assume the system to be in the presence of a strong coherent pump at frequency $\omega_p = \omega_c - \Delta$ which is used to enhance the weak optomechanical interaction
\begin{equation}
	H\s{pump} = i \mathcal E ( a_c^\da e^{ - i \omega_p t } - a_c e^{ i \omega_p t } ).
\end{equation}
The strong coherent pump creates a mean classical displacement of both the cavity mode and the mechanical mode, so that each mode experiences weak quantum fluctuations around those mean displacements.
We proceed by linearizing the dynamics around the classical displacements and consider the quantum evolution of the fluctuating parts.
Thereby in a rotating frame given by $H\s{rf}= \omega_p a_c^\da a_c + \omega_m a_m^\da a_m$ the linearized Hamiltonian takes the form (to simplify the equations we keep notation $a_i$ for only the fluctuating parts)
\begin{equation}
	H = \Delta a_c^\da a_c - g ( a_c^\da + a_c )( a_m^\da e^{ i \omega_m t } + a_m e^{ - i \omega_m t } ),
\end{equation}
where $g = g_0 \sqrt{\avg{ N\s{ph}}}$ is the coupling rate enhanced by the mean number of intracavity photons $\avg{ N\s{ph}}$.
This is the standard Hamiltonian of the linearized optomechanical interaction.
It has been successfully demonstrated in experiments to correctly describe the dynamics of the optomechanical systems in different regimes set by the detuning $\Delta$, e.g., in~\cite{groblacher_observation_2009,verhagen_quantum-coherent_2012,chan_laser_2011,*teufel_sideband_2011,palomaki_entangling_2013,millen_cavity_2015}.

In this manuscript we investigate the dynamics of an optomechanical system tuned to the blue sideband ($\Delta = - \omega_m$).
To derive the effective Hamiltonian for this regime, we first go to the rotating frame defined by $H\s{rf}' = \Delta a_c^\da a_c$, and then apply the rotating wave approximation to cancel the terms rapidly counter-rotating at $2 \omega_m$.
The latter requires the mechanical frequency $\omega_m$ be the dominant rate in the system, which practically is equivalent to the condition of the cavity decay $\kappa$ being much smaller than the mechanical frequency ($ \kappa \ll \omega_m$, so-called resolved sideband condition).
Then,
\begin{equation}
	\label{eq:ham:tms}
	H = - g ( a_c^\da  a_m^\da + a_c a_m ).
\end{equation}
This interaction is known to entangle the parties even in presence of large amounts of noise~\cite{filip_robust_2013,rakhubovsky_robust_2015}, so it is a natural choice to use this interaction for creating an entangled state of light and mechanics.
Unfortunately there is an inherent instability associated with this type of interaction in optomechanical systems~\cite{braginsky_notitle_1964,*braginsky_ponderomotive_1967}.
Tuned on the right slope of the cavity resonance curve (blue sideband), the coherent pump creates an effective spring for the mechanical mode.
Simultaneously, a negative damping (amplification) is created.
This negative damping might overwhelm the inherently low intrinsic damping of a high-Q mechanical system, thus rendering it unstable and making the steady state of the system inaccessible.
A natural way to compensate for this is to avoid steady states, and to shift attention to pulses~\cite{vanner_pulsed_2011,hofer_quantum_2011,vanner_cooling-by-measurement_2013}.

Using the Hamiltonian~\eqref{eq:ham:tms}, we can write the quantum Langevin equations introducing the damping and noise terms.
Since the Hamiltonian of the linearized optomechanical interaction is at most quadratic in the bosonic operators, the equations of motion are linear.
These equations can be written in the matrix form~\cite{genes_quantum_2009}
\begin{equation}
	\label{eq:vec:langevin}
	\dot{ \vec u } = \mat A \vec u + \sqrt{ 2 \mat K }\vec n,
\end{equation}
where $\vec u = (X_c , Y_c , X_m , Y_m)\up{T}$ is the vector of unknowns and $\vec n = (X\up{in}, Y\up{in}, X\up{th}, Y\up{th})\up{T}$ is the vector of the quadratures of the input quantum noises, the superscript $\up{T}$ denotes transposition.
The quadratures $X\up{in}$ and $Y\up{in}$ describe the input vacuum field of optical fluctuations, and $X\up{th}$ and $Y\up{th}$ are the quadratures of the thermal Langevin force acting upon the mechanical mode.
We use the convention $X_j = a_j + a_j^\da , Y_j = ( a_j - a_j^\da)/i$ for $j =c,m$, so that $\comm{X_j}{Y_j} = 2i$.
The operators of the environment obey $\comm{X^k (t)}{Y^k (t')} = 2 i \delta (t - t')$, for $k = \text{in, th}$.
Also we assume they obey the standard Markovian autocorrelations~\cite{giovannetti_phase-noise_2001}, e.g., $\avg{ X\up{in} (t) X\up{in} (t')} = \delta ( t - t ')$ and $\avg{ X\up{th}(t) X\up{th} (t')} = ( 2 n\s{th} + 1 ) \delta (t - t')$ with $n\s{th}$ being the average occupation of the bath.
Furthermore, $\mat K = \diag( \kappa , \kappa , \gamma /2 , \gamma/2 )$ is the diagonal matrix composed of optical damping rate $\kappa$ and the mechanical viscous damping rate $\gamma$, and $\mat A$ is the so-called drift matrix.
For the case of the two-mode squeezing interaction, the drift matrix has the form
\begin{equation}
	\mat A =
	\begin{pmatrix}
		- \kappa & 0        & g            & 0            \\
		0        & - \kappa & 0            & -g           \\
		g        & 0        & - \gamma / 2 & 0            \\
		0        & -g       & 0            & - \gamma / 2 \\
	\end{pmatrix}
\end{equation}
Here and throughout the article we assume $\avg{N\s{ph}}$ (and thereby $g$) independent of time.
This is a valid approximation for a top-hat drive with constant pump strength equal $\mathcal E (t) = \mathcal E$ within the pulse duration $0 \leq t \leq \tau$ and zero otherwise.
Another requirement is that the pulse duration is longer than the cavity decay time $\kappa \tau \gg 1$, for example as in~\cite{palomaki_entangling_2013}.

\cref{eq:vec:langevin} has the formal solution
\begin{equation}
	\label{eq:sol:int}
	\vec u (t) = \mat M ( t ) \vec u (0) + \int _0 ^t  \dd s \mat M ( t - s ) \sqrt{ 2 \mat K } \vec n ( s ),
\end{equation}
where
\begin{equation}
	\mat M ( s ) = \exp [ \mat A s ].
\end{equation}
Importantly, for a bipartite system, an analytical form of $\mat M$ can always be computed however in general it is too cumbersome to be reported here.
In the general case of a time-dependent pump strength,~\cref{eq:vec:langevin} should be solved numerically.

\section{Optimized temporal modes of light} 
\label{sec:temporal_modes_of_light}

The solution~\eqref{eq:sol:int} is sufficient to analyze the optomechanical dynamics, particularly the transfer of quantum information from the incident light to the mechanical mode.
Among other terms, the expression for the mechanical mode contains a term corresponding to the input optical quantum fluctuations serving as the signal
\begin{multline}
	X_m (\tau) = \dots + \int_0^\tau \dd s \mat M_{31} (\tau -s ) \sqrt{ 2 \kappa } X\up{in} (s)
	\\
	=
	\dots + \sqrt{ \cG - 1 } \Mode X\up{in},
\end{multline}
where we have designated the quadratures of the input optical temporal mode and the optomechanical amplification gain $\cG$ as
\begin{gather}
	(\Mode X\up{in} , \Mode Y\up{in} )\up{T}  \equiv \int_0^\tau \dd s f\up{in} (s) ( X\up{in} (s) , Y\up{in} (s) )\up{T},
	\\
	f\up{in} (s)                              = \sqrt{ \frac{ 2 \kappa }{ \cG - 1} } \mat M_{31} (\tau - s),
	\\
	\cG                                       =  1 + 2 \kappa  \int _0^\tau  \dd s \mat M_{31}^2 ( s ).
\end{gather}
The definition of $\Mode Y\up{in}$ features the same $f\up{in}$ as the one for $\Mode X\up{in}$ owing to the symmetries in $\mat A$ that secure $\mat M_{31} (s) = - \mat M_{42} (s)$.
This way we select a single temporal mode--- the optimally coupled one--- from the continuum of the temporal modes of the input fluctuations.
By definition, this mode has proper commutation relations $\comm{ \Mode X\up{in} }{ \Mode Y\up{in} } = 2i$, because $\int_0^\tau \dd s ( f\up{in} (s) )^2 = 1.$

To complete the consideration of the system, we have to supplement the equations above with an input-output relation in the form
\begin{equation}
	\label{eq:input-output}
	\vec v\up{out} (t) = - \tilde { \vec n}\up{in} (t) + \sqrt{ 2 \kappa } \tilde{ \vec u} (t),
\end{equation}
where $\vec v\up{out} = ( X\up{out} , Y\up{out} )\up{T}$, and the tilde denotes that we take the only first two entries of the vector, e.g. $\tilde{ \vec u } = ( \vec u _1 , \vec u_2 )\up{T}$.

In a similar fashion with the solution for the intracavity quadratures, we conclude that the expression for the leaking field has a term
\begin{equation}
	X\up{out} (t) = \dots + \mat M_{13} (t) \sqrt{ 2 \kappa } X_m (0).
\end{equation}
Therefore, the proper temporal mode of the leaking field, which is optimally coupled to the mechanical mode, is defined by the profile of the interaction $\mat M_{13} (t)$.
We thus define the quadratures of the output mode in full analogy with the definition for the input mode:
\begin{align}
	\label{eq:def_out_mode}
	(\Mode X\up{out} , \Mode Y\up{out} )\up{T} & \equiv \int_0^\tau \dd s f\up{out} (s) ( X\up{out} (s) , Y\up{out} (s) )\up{T},
	\\
	f\up{out} (s)                              & = \sqrt{ \frac{ 2 \kappa }{ \cG - 1} } \mat M_{13} (s),
\end{align}
Again, owing to the symmetries in $\mat A$, $\mat M\up{T} = \mat M$ and therefore
\begin{equation}
	\int_0^\tau \dd s \mat M_{13}^2 ( s ) = \int_0^\tau \dd s \mat M_{31}^2 (s) = \frac{ \cG - 1 }{ 2 \kappa }.
\end{equation}

Defined by~\eqref{eq:def_out_mode}, the output mode obeys proper commutations $\comm{ \Mode X\up{out}}{ \Mode Y\up{out}} = 2 i$, and carries maximum information regarding the initial mechanical state.
Note that if, instead, a different temporal mode is measured it will carry only a certain fraction of the information, with this fraction defined by the temporal overlap of the profiles of the modes.

In the general case, the elements of $\mat M$ and, therefore, the temporal shapes of the optimal modes of the radiation, have a complicated dependence on time.
In certain important cases, however, there exist simple approximations to the exact temporal shapes.
Particularly, for the case of a two-mode squeezing optomechanical interaction with Hamiltonian~\eqref{eq:ham:tms},  the corresponding functions take the form
\begin{equation}
	f\up{out} (s) \propto \mat M_{13}(t)
	= \frac g \lambda \left[ e^{ - \tfrac t2 ( \kappa + \frac \gamma 2 - \lambda  ) } - e^{ - \tfrac t 2 ( \kappa + \frac \gamma 2 + \lambda ) } \right],
\end{equation}
with $\lambda = \sqrt{ ( \kappa - \gamma /2 )^2 + 4 g^2 }$.
In the case of wide-band cavity $\kappa \gg g , \gamma$ the intracavity mode can almost instantaneously react to the other influences, and therefore can be adiabatically eliminated.
Thereby, in this \emph{adiabatic regime} the expression for the leaking mode shape can be simplified and transforms into
\begin{equation}
	f\up{out}\s{ad} (t) \propto e^{ G t },
\end{equation}
where we define the optomechanical amplification rate $G = g^2 / \kappa$.
In a similar fashion one can process the temporal shape of the incident pulse to obtain the expressions of the input and output modes' temporal shape functions as follows
\begin{equation}
	\label{eq:exponential_pulses}
	f\s{ad}\up{in} (t)  = \sqrt{ \frac{ 2 G }{ 1 - e^{  - 2 G \tau }}} e^{ - G t },
	\quad
	f\s{ad}\up{out} (t) = \sqrt{ \frac{ 2 G }{ e^{  2 G \tau } - 1}}   e^{ G t }  ,
\end{equation}
The input-output transformations for the optical and the mechanical mode then read (for rather long pulses $\kappa \tau \gg 1 $ and ignoring the mechanical decoherence for now)
\begin{subequations}
	\label{eq:ioblue:boson}
	\begin{align}
		\Mode A\up{out} & = \sqrt{ \cG } \Mode A\up{in} + \sqrt{ \cG - 1 } a_m^\da (0),
		\\
		a_m (\tau)      & = \sqrt{ \cG } a_m (0) + \sqrt{ \cG - 1 } \Mode A\up{in,}{}^\da.
	\end{align}
\end{subequations}
The Bogoliubov transformations above correspond to the two-mode squeezing of the optical and the mechanical modes with the amplification coefficient $\cG = e^{ 2 G \tau }$.
For the sake of compactness of notation we have introduced the bosonic operators $\Mode A^k = \frac 12 ( \Mode X^k + i \Mode Y^k ),$ for $k = \text{in, out}$.

\section{Conditional squeezing in the adiabatic regime} 
\label{sec:conditional_squeezing_in_the_adiabatic_regime}

The two-mode squeezing interaction is known for its ability to entangle the modes and produce correlations sufficient to achieve conditional squeezing (CS)~\cite{filip_robust_2013,rakhubovsky_robust_2015}.
The latter effect manifests itself as a projection of one of the two correlated modes onto a state which has the variance of one of the quadratures below the shot-noise level.
In this section we introduce the formal definition of this effect and reiterate the sufficient conditions for the observation of CS in an experiment.

The state of the optomechanical system after the interaction is characterized by the vector $\vec r = (\Mode X\up{out} , \Mode Y\up{out} , X_m (\tau) , Y_m (\tau))\up{T}$.
Since the system undergoes linear dynamics described by~\eqref{eq:vec:langevin} with the initial state being Gaussian, the final state of the system remains Gaussian and can therefore be completely described by a vector of means $\avg{ \vec r }$ and the covariance matrix (CM) $\mat V$.
The latter is defined as a matrix with elements
\begin{equation}
	\label{eq:covm:optomech}
	\mat V_{ij } = \smtr{ \vec{ r } }{ \vec{ r} }_{ij } \equiv
	\tfrac 12 \avg{ \acomm{( \vec r_i  - \avg{ \vec r}_i )}{( \vec r_j  - \avg{ \vec r}_j )} } ,
\end{equation}
where $\acomm a b \equiv a b + b a$ is the anticommutator, and the averaging is performed in the quantum mechanical sense $\avg{ a } \equiv \Tr ( \rho a ).$

Regrouping for convenience the elements of the vector $\vec r$ into $\vec r\up{out} = (\Mode X\up{out} , \Mode Y\up{out} )\up{T}$ and $\vec r_m = (X_m , Y_m)\up{T}$, we can write the CM in the block form
\begin{equation}
	\label{eq:cm:block}
	\mat V =
	\begin{pmatrix}
		\mat V\up{out}       & \mat V^c
		\\
		\mat V^{c, }{}\up{T} & \mat V^m
	\end{pmatrix}
\end{equation}
with
\begin{equation}
	\notag
	\mat V\up{out} = \smtr{ \vec r\up{out} }{ \vec r\up{out} } ,
	\mat V^m = \smtr{ \vec r_m }{ \vec r_m },
	\mat V^c =\smtr{ \vec r\up{out} }{ \vec r_m }.
\end{equation}
The blocks on the main diagonal show the autocorrelations of the leaking pulse and the mechanical mode respectively, and the off-diagonal block show the cross-correlations between the two.

Upon homodyne detection of an amplitude quadrature ($\Mode X\up{out}$) of the leaking pulse, the mechanical mode is projected onto a Gaussian state with CM $\mat V^m{}'$ given by~\cite{weedbrook_gaussian_2012}
\begin{equation}
	\label{eq:homodyne:covm}
	\mat V^m{}' = \mat V^m - \left[ \mat V\up{out}_{11 }\right]^{-1} \mat V^c{}\up{,T} \mat P \mat V^c,
\end{equation}
with $\mat P = \diag (1,0)$.
For the generalization to the case of an arbitrary quadrature homodyne measurement see Appendix~\ref{sec:covariance_matrix_after_homodyne_measurement}.
The state is squeezed if the smaller eigenvalue $\sigma\s{cond}$ (for which we use the term \emph{conditional variance}) of $\mat V^m{}'$ is below the uncertainty of the vacuum $\sigma\s{vac}$.
For our choice of the commutation relations $\comm{ X_m }{ Y_m } = 2 i$,  $\sigma\s{vac} = 1$.
In the simple case of a diagonal $\mat V^m{}'$ the matrix elements on the principal diagonal have the meaning of uncertainties of the quadratures.
The conditional variance is therefore equal to the smaller diagonal element and has the very illustrative meaning of the uncertainty of the squeezed quadrature.

In the case of the two-mode squeezing interaction, described by~\eqref{eq:ioblue:boson}, the mechanical mode is projected by the homodyne detection of the amplitude quadrature of the leaking pulse on the state with the covariance matrix~\cite{filip_robust_2013}
\begin{equation}
	\label{eq:covmat:adiabat}
	\mat V^m{}' = \diag \left( \frac{ 2 n_0 + 1 }{ \cG +  ( \cG - 1 ) ( 2 n_0 + 1 ) },  2 \cG ( n_0 + 1 ) - 1 \right),
\end{equation}
with $n_0$ being the initial occupation of the mechanical mode.
In the experimentally relevant limit of high occupation, the covariance matrix simplifies to
\begin{equation}
	\label{eq:covmat:adiabat:highT}
	\mat V^m{}' \Big\vert_{n_0 \gg 1} = \diag\left( \frac{ 1 }{ \cG - 1 }, 2 \cG n _0 \right),
\end{equation}
which is squeezed regardless of the initial occupation provided $\cG > 2$.
Importantly, the squeezing occurs upon measurement of an \emph{arbitrary} quadrature which is a manifestation of the phase insensitivity of the two-mode squeezing interaction.
That is, regardless of which quadrature of light is measured, the eigenvalues of the covariance matrix $\mat V^m{}'$ remain the same, although the eigenvectors might be rotated.
Moreover, the final state of the mechanical mode is displaced in the phase space dependent on the outcome of the measurement of the optical mode.
Such a displacement however does not affect the nonclassicality.

\section{Realistic conditional squeezing of a levitated nanoparticle} 
\label{sec:conditional_squeezing_of_a_levitated_nanoparticle}

\begin{figure}[htb]
	\centering
	\includegraphics[width=.99\linewidth]{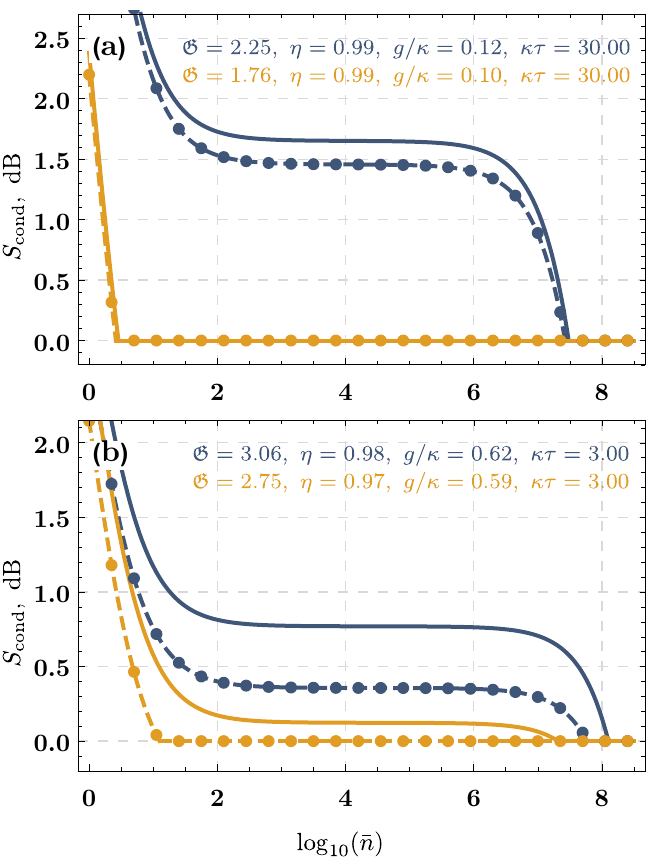}
	\caption{Conditional squeezing of a levitated particle as a function of temperature.
		We consider the mechanical mode to be in equilibrium with the environment before the pulsed protocol, so $n_0 = n\s{th} = \bar n$.
		Both panels make use of full solution of~\cref{eq:vec:langevin} without further approximations.
		(a)~--- close to the adiabatic regime, (b)~--- beyond the adiabatic regime.
		Different colors correspond to different choices of coupling strength $g$ and the pulse duration $\tau$, leading to different values of effective gain $\cG$ and loss $\eta$.
		The plateau in squeezing magnitudes, predicted by the adiabatic regime, remains in the full solution.
		Solid and dashed lines correspond to an ideal detection of, respectively, the optimal~\eqref{eq:def_out_mode} and the exponential~\eqref{eq:exponential_pulses} mode of the leaking pulse.
		The lines with markers overlapping with the dashed lines correspond to the detection of the optimal mode that experienced a loss with transmittance equivalent to that provided by the overlap of the optimal and the exponential modes envelope (see~\eqref{eq:eta_overlap_def}).
	}
	\label{fig:cond_squeezing}
\end{figure}

As can be seen from the previous section, under idealized conditions, the blue-detuned optomechanical interaction produces the thermal equivalent of a two-mode squeezed state of the leaking field and the levitated particle.
In this section we evaluate the effect of imperfections in the system and analyze the potential to achieve conditional squeezing in state-of-the art experiments.

The idealized solution~\eqref{eq:ioblue:boson} analyzed in the previous section does not include mechanical decoherence, involves adiabatic elimination of the cavity mode and ignores optical losses.
The first two effects are taken into account by considering the full solution of~\eqref{eq:vec:langevin}; the machinery for this evaluation is described in Appendix~\ref{app:estimating_the_covariance_matrix}.
The optical loss enters at different stages of the protocol and in particular it manifests itself in nonunit photon escape and photodetection efficiencies, losses in the mirrors and in propagation etc.
Advantageously, all these effects can be treated as a linear admixture of vacuum to the output signal (see Fig.~\ref{fig:fig0:scheme}).
Thereby, the optical loss can be taken into account by modifying the input-output relation~\eqref{eq:input-output} as
\begin{multline}
	\label{eq:io:loss}
	\vec v\up{out} (t) = \sqrt{ \eta } \left[
		- \tilde { \vec n}\up{in} (t) + \sqrt{ 2 \kappa } \tilde{ \vec u} (t)
		\right]
	\\
	+ \sqrt{ 1 - \eta } \vec v\up{vac} (t),
\end{multline}
with $0 \leq \eta \leq 1$ being the total transmittance associated with all the loss sources combined ($\eta=1$ corresponding to the lossless case), and $\vec v\up{vac} = (\Mode X\up{vac} , \Mode Y\up{vac})\up{T}$ describing the joint vacuum mode of light.
Following the modification described by~\eqref{eq:io:loss}, the covariance matrix $\mat V$ becomes
\begin{equation}
	\mat V \to
	\begin{pmatrix}
		\eta \mat V\up{out}          & \sqrt{ \eta } \mat V^c
		\\
		\sqrt \eta \mat V^c{}\up{,T} & \mat V^m
	\end{pmatrix}
\end{equation}
Note that we do not have to consider the losses associated with the input light, since it is already in vacuum state, thereby an admixture of vacuum fluctuations to it does not modify the statistics of the input light.

The results of our simulations can be seen at Fig.~\ref{fig:cond_squeezing}.
We plot the conditional squeezing in dB as a function of the initial occupation of the levitated nanoparticle which we assume to be in equilibrium with its environment (so that $n_0 = n\s{th} = \bar n$).
The plotted quantity is therefore
\begin{equation}
	S\s{cond} = \max \left( 0 , - 20 \log_{10} \left[ \sigma\s{cond} (\bar n) \right] \right).
\end{equation}

Our analysis shows that in the regime of the parameters satisfying the requirements of the adiabatic interaction ($\kappa \gg g , \tau^{-1}, \gamma$  and $\gamma n\s{th} \tau \ll 1$) the full solution exhibits good correspondence with the approximate adiabatic one.
As the occupation increases from the ground state, the magnitude of squeezing decreases until it reaches a plateau that spans over a few orders of magnitude of the occupation number values.
The plateau is as well predicted by the adiabatic regime (cf.~\cref{eq:covmat:adiabat,eq:covmat:adiabat:highT}).
In accordance with this prediction, the magnitude of squeezing is determined solely by the amplification gain $\cG$ albeit the dependence is more complicated than in the adiabatic regime.
Once the gain is above a certain threshold, the CS is possible for a wide range of temperatures, impossible for gain values below threshold (compare blue [darker] and yellow [lighter] lines in Fig.~\ref{fig:cond_squeezing}~(a)).
The CS occurs regardless of which quadrature of the leaking field is measured, which is consistent with the phase insensitivity of two-mode squeezing.

The most important limiting factor in conditionally squeezing the mechanics is the nonzero initial occupation $n_0$ of the levitated particle and the occupation $n\s{th}$ of its bath.
In Fig.~\ref{fig:cond_squeezing} we assume that the particle is initially at equilibrium with the surroundings, so $n_0 = n\s{th}$.
The two occupations though have different effects on the possibility of the CS.
For low values of the initial mechanical occupation $n_0$, before the plateau, the value of $n_0$ sets the upper boundary for the magnitude of the CS of the mechanical oscillator in accordance with~\eqref{eq:covmat:adiabat}.
This occupation alone however does not impose limits on the CS, that is, in the case of zero occupation of the bath $n\s{th} = 0$, CS is possible for an arbitrary $n_0$ provided that the optomechanical amplification gain $\cG$ is above a certain threshold.
Additionally the occupation of the bath $n\s{th}$ does not influence the magnitude of the CS as long as it is below a certain value related to the rethermalization time $n\s{th} \lesssim n\up{crit} \equiv (\gamma \tau)^{-1}$.
Above this value of $n\s{th}$ the CS is impossible regardless of the initial occupation $n_0$.
We would like to emphasize that this means that if for a given temperature of the bath the CS happens to be impossible, precooling alone does not salvage the situation.

Intuitively, in order to push the boundary of the temperatures allowing the CS up, one would shorten the pulse duration $\tau$ which means a shorter interaction with the bath and therefore a smaller amount of noise entering the mechanical mode.
Indeed this helps however the appropriate values of $\tau$ are limited from below by the inverse cavity linewidth: $\tau \gtrsim \kappa^{-1}$, apparently because shorter pulses are unable to properly enter and leave the cavity.
This, combined with the requirement of a resolved side-band, $\kappa \ll \omega_m$, sets a limit on the occupations of the bath, allowing the CS.
The upper boundary for the phonon number is $n\up{crit} \lesssim  \omega_m / \gamma  =  Q_m$.

Decreasing the pulse duration $\tau$ causes a decrease of the amplification gain $\cG$ and thereby requires an increase in the optomechanical interaction rate $g$ in order to compensate it.
The latter can be increased by increasing the intracavity power and subsequently the average number of the intracavity photons $\avg{N\s{ph}}$.
Decreasing $\tau$, and increasing $g$ such that it becomes comparable with $\kappa$ leads us farther from the adiabatic regime, which means that the optimal temporal mode of the leaking field increasingly differs from the simple exponential form~\eqref{eq:exponential_pulses}.
The analysis presented in Fig.~\ref{fig:cond_squeezing} shows that the detection of an approximate exponential mode~\eqref{eq:exponential_pulses} (dashed lines) is indistinguishable from the detection of the optimal mode~\eqref{eq:def_out_mode} preceded by an optical loss (lines with markers).
The associated transmittance $\eta$ of the loss is exactly equivalent to the overlap of the temporal profiles of the modes:
\begin{equation}
	\label{eq:eta_overlap_def}
	\eta\s{mm} = \int_0^\tau \dd s f\up{out} (s) f\up{out}\s{ad} (s),
\end{equation}
where $f\up{out} (s) \propto \mat M_{13} (s)$ is defined by the full solution of the equations of motion, and $f\up{out}\s{ad} (s)$, by the adiabatic solution~\eqref{eq:exponential_pulses}.
Both these functions are illustrated in the inset of Fig.~\ref{fig:fig0:scheme}.
From the comparison of the magnitudes of CS attainable by the lossless and lossy detection it is evident that the optical loss can be a critical limiting factor to the extent that it renders the CS impossible at the temperatures at which a lossless detection would allow (see yellow [lighter] lines in~\cref{fig:cond_squeezing}~(b)).
Note however that the deviation from the adiabatic regime, which invalidates the elimination of the cavity mode, does not cause any loss itself.
The contribution of the cavity mode, that can not be eliminated anymore, is the redistribution of the quantum information between the different temporal modes of the continuum of the modes of the leaking radiation.
This effect can be compensated totally by optimization of the profile of the detected mode.

\section{Conclusion} 
\label{sec:conclusion_and_outlook}

We have considered the perspective of conditional squeezing of a levitated nanoparticle by a pulsed protocol including entangling the particle with the control field, and then homodyne detection of the field.
We have shown that within the resolved sideband regime, on the timescales shorter than the rethermalization time ($\tau \ll [ \gamma n\s{th} ]^{-1}$) the interaction in the system is ultimately approaching the two-mode squeezing and can allow the conditional squeezing regardless of the initial occupation of the levitated particle.
Importantly, cooling of the particle while remaining at the same temperature of the bath is unable to enhance the squeezing.

To take advantage of the created quantum correlations one has to detect a proper temporal mode of the leaking radiation.
The profile of this temporal mode is set by the optomechanical interaction, and in the regime far from that typically considered adiabatic (where the condition $\kappa \gg g$ does not hold anymore) this profile differs from the standard exponential one.
Detection of a different mode causes loss of the correlations which might prohibit the squeezing of the mechanics.

Our analysis is carried out in the dimensionless variables which allows it to be translated to an arbitrary optomechanical system capable of working in the resolved sideband regime $\kappa \ll \omega_m$.
We find the domain of levitated nanoparticles to be a promising candidate for implementation, as it allows good control of the experiment, outstanding isolation from the environment (and thereby exceptionally high mechanical $Q$-factors) and relatively strong optomechanical coupling $g$ compared to the standard bulk systems.
To produce~\cref{fig:cond_squeezing} we used the parameters reported in~\cite{delic_cavity_2018}.
In that setup $\omega_m = 2 \pi \times  180$~kHz, $\kappa = 2 \pi \times 96.5$~kHz, and $g \leq 2 \pi \times {60}$~kHz which corresponds to the maximal value of the ratio $g/\kappa = 0.62$.
We also used for the estimations the value $\gamma / \kappa = 2.8\times 10^{-10}$.
Moreover, the conditional squeezing is already possible at moderate coupling rates as is shown in~\cref{fig:cond_squeezing}~(a).
Rates of this order have been reported in a number of optomechanical experiments with bulk mechanical oscillators (see e.g.,~\cite{teufel_sideband_2011,chan_laser_2011}).
The proposed protocol is therefore definitely feasible in a state of the art experiment.

Detection of the mechanical squeezing requires a detection of mechanical displacement with sub-shot-noise precision.
This can be done by different means, e.g., either by a back-action-evading operation~\cite{clerk_back-action_2008,*suh_mechanically_2014,*shomroni_optical_2018}, a squeezing-enhanced~\cite{kerdoncuff_squeezing-enhanced_2015} or a variational~\cite{mason_continuous_2018} measurement, or by swapping the mechanical state with the state of a red-detuned pulse~\cite{hofer_quantum_2011,palomaki_entangling_2013,filip_transfer_2015} and subsequent optical tomography.
Evaluation of this task however goes beyond the scope of the current manuscript.

We have thus shown a possibility to create a motional nonclassical state of a levitated particle by a sequence of optomechanical interaction and an optical detection.
This possibility does not require cooling the particle to the ground state and is within experimental reach.
Squeezing of a mechanical oscillator below the shot noise level paves the way to high-precision measurements and tests of fundamental science, such as quantum mechanics~\cite{kaltenbaek_macroscopic_2016} and thermodynamics~\cite{gieseler_levitated_2018}.

\section*{Acknowledgments} 
\label{sec:acknowledgments}

The authors are grateful to Nikolai Kiesel and Uro{\v s} Deli{\' c} for fruitful discussions.
The authors acknowledge the support of the project GB14-36681G of the Czech Science Foundation.
The authors have received national funding from the MEYS under grant agreement No.
731473 and from the QUANTERA ERA-NET cofund in quantum technologies implemented within the European Union's Horizon 2020 Programme (project TheBlinQC).
A.A.R. and D.W.M. acknowledge the Development Project of Faculty of Science, Palack{\' y} University.
A.A.R. also acknowledges support by the project LTC17086 of INTER-EXCELLENCE program of the Czech Ministry of Education.

\appendix

\section{Estimating the covariance matrix} 
\label{app:estimating_the_covariance_matrix}

In this section we outline the process of evaluation of the covariance matrix~\eqref{eq:cm:block}, particularly, its first diagonal block $\smtr{\vec r\up{out}}{\vec r\up{out}}$ showing the autocorrelations of the leaking pulse.
These considerations can be extended for the evaluation of the other blocks in a somewhat obvious way.

We have for $\vec r\up{out}$ the expression (for $i = 1,2$.
We also assume summation over repeating indices, e.g., $a_j b_j$ should read $\sum_{j=1}^4 a_j b_j$)
\begin{widetext}
	\begin{multline}
		\vec r\up{out}_i =
		\int_0^\tau \dd s f (s) \Big[  - \vec r\up{in}_i (s) + \sqrt{ 2 \kappa } \mat M_{ij } (s ) \vec u_j (0)
			+ \sqrt{ 2 \kappa } \int_0^s \dd{s'} \mat M_{ij } (s - s') \sqrt{ 2 \mat K_{jk } } \vec r\up{in}_k (s') \Big]
		\\
		= - \int_0^\tau \dd s f(s) \vec r_i\up{in} (s)
		+ \sqrt{ 2 \kappa } \vec u_j (0) \int_0^\tau \dd s f(s) \mat M_{ij } (s )
		+ \sqrt{ 2 \kappa} \int_0^\tau \dd s \vec r\up{in}_k (s) \int_s^\tau \dd{s'} f(s') \mat M_{ij } (s' - s) \sqrt{ 2 \mat K_{jk } }.
	\end{multline}

	We then insert this into the definition of the CM and use the linearity of the operation of covariance.
	Moreover, we take advantage of the fact that the initial values of quadratures $\vec u (0)$ are uncorrelated with the quantum fluctuations $\vec r\up{in}$ (the Langevin forces in~\eqref{eq:vec:langevin}):
	\begin{multline}
		\smtr{ \vec r\up{out} }{ \vec r\up{out} }_{ii' } =
		2 \kappa \smtr{ \vec u_j (0) }{ \vec u_{j'} (0) } \int_0^\tau \dd s \dd t f(s) f(t) \mat M_{ij } (s) \mat M_{i'j' } (t) + \iint_0 ^ \tau \dd s \dd t f(s) f(t) \smtr{ \vec r_i\up{in} (s) }{ \vec r_{i' }\up{in} (t) }
		\\
		- 2 \sqrt{ 2 \kappa } \iint_0^\tau \dd s \dd t f (s) \smtr{ \vec r_i \up{in} (s) }{ \vec r_{k' }\up{in} (t) } \int_t^\tau \dd{t'} f(t') \mat M_{i'j'} (t' - t ) \sqrt{ 2 \mat K_{j' k'} }
		\\
		+ 4 \kappa \iint _0^\tau \dd s \dd t \smtr{ \vec r\up{in}_k (s) }{ \vec r\up{in}_{k' } (t) } \int_s^\tau \dd{s'} \int_t^\tau \dd {t'} f (s') f(t') \mat M_{ij } (s' - s ) \mat M_{i'j' } (t' - t ) \sqrt{ \mat K_{jk }} \sqrt{ \mat K_{j' k'}}.
	\end{multline}
	The statistics of the initial state and of the Langevin forces are known.
	Typically in experiment the mechanical mode is initially in the thermal state with occupation $n_0$, and the cavity mode is initially in vacuum.
	Similarly, the mechanical bath is in the thermal state with the occupation $n\s{th}$ and the input optical fluctuations are in vacuum.
	Therefore,
	\begin{gather}
		\smtr{ \vec u_i (0) }{ \vec u_j (0) } = \sigma_{ij }^{(0)} = \diag[ 1 , 1 , 2 n_0 + 1 , 2 n_0 + 1 ];
		\\
		\smtr{ \vec r_i\up{in} (t) }{ \vec r_j\up{in} (t') } = \sigma_{ij }\up{in} \delta ( t - t' ) = \diag [ 1 , 1 , 2 n\s{th} + 1 , 2 n\s{th} + 1 ] \delta ( t - t' ).
	\end{gather}
	We have then (for $i,i' = 1,2$)
	\begin{multline}
		\smtr{ \vec r\up{out} }{ \vec r\up{out} }_{ii' } =
		\sigma\up{in}_{ii'}
		+ 2 \kappa \sigma_{jj'} ^{(0)} \int_0^\tau \dd s \dd t f(s) f(t) \mat M_{ij } (s) \mat M_{i'j' } (t)
		- 2 \sqrt{ 2 \kappa } \sigma\up{in}_{ik' } \int_0^\tau \dd t f (t) \int_t^\tau \dd{t'} f(t') \mat M_{i'j'} (t' - t ) \sqrt{ 2 \mat K_{j' k'} }
		\\
		+ 4 \kappa \sigma\up{in}_{kk'} \int _0^\tau \dd t \iint_t^\tau \dd{s'} \dd {t'} f (s') f(t') \mat M_{ij } (s' - t ) \mat M_{i'j' } (t' - t ) \sqrt{ \mat K_{jk }} \sqrt{ \mat K_{j' k'}}.
	\end{multline}
\end{widetext}
The computation of the covariance matrix is, thereby, reduced to ordinary integration.
For the case of the two-mode squeezing interaction~\eqref{eq:vec:langevin} it can be performed analytically.

\section{Covariance matrix after a homodyne measurement of an arbitrary quadrature} 
\label{sec:covariance_matrix_after_homodyne_measurement}

In this section, following~\cite{eisert_distilling_2002,*fiurasek_gaussian_2002,*serafini_quantum_2017} we present an expression for the conditional covariance matrix of the mechanical mode after a homodyne measurement of an arbitrary quadrature of the leaking light.

Consider an optomechanical system in a Gaussian state described by a CM of the form~\eqref{eq:cm:block}.
A measurement of the optical mode that projects it on a pure state with CM~$\mat D$ simultaneously projects the mechanical mode on a state with CM that reads
\begin{equation}
	\label{eq:gaussian:measurement}
	\mat V^{m'} = \mat V^m  - \mat V^{c,\text{T}} ( \mat V\up{out} + \mat D )^{-1} \mat V^c.
\end{equation}
For a homodyne measurement of the amplitude quadrature $X$,
\begin{equation}
	\mat D = \lim_{d \to 0} \diag [ d , d^{-1} ],
\end{equation}
and the inverse in~\cref{eq:gaussian:measurement} is a pseudoinverse.
A proper simplification yields~\cref{eq:homodyne:covm}.

To generalize the~\cref{eq:gaussian:measurement} to the measurement of an arbitrary quadrature $X_\theta = X \cos \theta + Y \sin \theta$ we note that such a measurement projects the optical mode on a state with the covariance matrix
\begin{equation}
	\mat D_\theta = \lim_{d \to 0 } \mat R_2 ( - \theta) \mat D \mat R_2 ( - \theta )\up{T},
\end{equation}
where $\mat R_2 (\theta)$ is a $2\times 2 $ rotation matrix
\begin{equation}
	\mat R_2 (\theta) =
	\begin{pmatrix}
		\cos \theta   & \sin \theta
		\\
		- \sin \theta & \cos \theta
	\end{pmatrix}.
\end{equation}
Using $\mat D_\theta$ in~\cref{eq:gaussian:measurement} instead of $\mat D$ gives the required covariance matrix.

A less technically demanding way to obtain the required CM is to note that a homodyne detection of the quadrature $X_\theta$ is equivalent to a homodyne detection of the amplitude quadrature in a rotated basis.
In this basis the CM of the optomechanical system reads
\begin{equation}
	\label{eq:rotated:cm}
	\mat V (\theta) = \mat R (\theta ) \mat V \mat R (\theta )\up{T},
\end{equation}
where $\mat V$ is given by~\cref{eq:cm:block}, and $\mat R$ performs rotation of only the basis of the optical mode:
\begin{equation}
	\mat R (\theta) = \mat R_2 (\theta) \oplus \diag [ 1 , 1 ].
\end{equation}
Applying the formalism of~\cref{eq:homodyne:covm} to $\mat V(\theta)$ gives the CM of the mechanical mode after the measurement.

\bibliography{levitated_ascii}

\begin{thebibliography}{79}%
\makeatletter
\providecommand \@ifxundefined [1]{%
 \@ifx{#1\undefined}
}%
\providecommand \@ifnum [1]{%
 \ifnum #1\expandafter \@firstoftwo
 \else \expandafter \@secondoftwo
 \fi
}%
\providecommand \@ifx [1]{%
 \ifx #1\expandafter \@firstoftwo
 \else \expandafter \@secondoftwo
 \fi
}%
\providecommand \natexlab [1]{#1}%
\providecommand \enquote  [1]{``#1''}%
\providecommand \bibnamefont  [1]{#1}%
\providecommand \bibfnamefont [1]{#1}%
\providecommand \citenamefont [1]{#1}%
\providecommand \href@noop [0]{\@secondoftwo}%
\providecommand \href [0]{\begingroup \@sanitize@url \@href}%
\providecommand \@href[1]{\@@startlink{#1}\@@href}%
\providecommand \@@href[1]{\endgroup#1\@@endlink}%
\providecommand \@sanitize@url [0]{\catcode `\\12\catcode `\$12\catcode
  `\&12\catcode `\#12\catcode `\^12\catcode `\_12\catcode `\%12\relax}%
\providecommand \@@startlink[1]{}%
\providecommand \@@endlink[0]{}%
\providecommand \url  [0]{\begingroup\@sanitize@url \@url }%
\providecommand \@url [1]{\endgroup\@href {#1}{\urlprefix }}%
\providecommand \urlprefix  [0]{URL }%
\providecommand \Eprint [0]{\href }%
\providecommand \doibase [0]{http://dx.doi.org/}%
\providecommand \selectlanguage [0]{\@gobble}%
\providecommand \bibinfo  [0]{\@secondoftwo}%
\providecommand \bibfield  [0]{\@secondoftwo}%
\providecommand \translation [1]{[#1]}%
\providecommand \BibitemOpen [0]{}%
\providecommand \bibitemStop [0]{}%
\providecommand \bibitemNoStop [0]{.\EOS\space}%
\providecommand \EOS [0]{\spacefactor3000\relax}%
\providecommand \BibitemShut  [1]{\csname bibitem#1\endcsname}%
\let\auto@bib@innerbib\@empty
\bibitem [{\citenamefont {Meystre}(2013)}]{meystre_short_2013}%
  \BibitemOpen
  \bibfield  {author} {\bibinfo {author} {\bibfnamefont {Pierre}\ \bibnamefont
  {Meystre}},\ }\bibfield  {title} {\enquote {\bibinfo {title} {A short walk
  through quantum optomechanics},}\ }\href {\doibase 10.1002/andp.201200226}
  {\bibfield  {journal} {\bibinfo  {journal} {Annalen der Physik}\ }\textbf
  {\bibinfo {volume} {525}},\ \bibinfo {pages} {215--233} (\bibinfo {year}
  {2013})}\BibitemShut {NoStop}%
\bibitem [{\citenamefont {Aspelmeyer}\ \emph {et~al.}(2014)\citenamefont
  {Aspelmeyer}, \citenamefont {Kippenberg},\ and\ \citenamefont
  {Marquardt}}]{aspelmeyer_cavity_2014}%
  \BibitemOpen
  \bibfield  {author} {\bibinfo {author} {\bibfnamefont {Markus}\ \bibnamefont
  {Aspelmeyer}}, \bibinfo {author} {\bibfnamefont {Tobias~J.}\ \bibnamefont
  {Kippenberg}}, \ and\ \bibinfo {author} {\bibfnamefont {Florian}\
  \bibnamefont {Marquardt}},\ }\bibfield  {title} {\enquote {\bibinfo {title}
  {Cavity optomechanics},}\ }\href {\doibase 10.1103/RevModPhys.86.1391}
  {\bibfield  {journal} {\bibinfo  {journal} {Reviews of Modern Physics}\
  }\textbf {\bibinfo {volume} {86}},\ \bibinfo {pages} {1391--1452} (\bibinfo
  {year} {2014})},\ \bibinfo {note} {arXiv: 1303.0733}\BibitemShut {NoStop}%
\bibitem [{\citenamefont {Khalili}\ and\ \citenamefont
  {Danilishin}(2016)}]{khalili_quantum_2016}%
  \BibitemOpen
  \bibfield  {author} {\bibinfo {author} {\bibfnamefont {Farid~Ya.}\
  \bibnamefont {Khalili}}\ and\ \bibinfo {author} {\bibfnamefont {Stefan~L.}\
  \bibnamefont {Danilishin}},\ }\bibfield  {title} {\enquote {\bibinfo {title}
  {Quantum {Optomechanics}},}\ }in\ \href@noop {} {\emph {\bibinfo {booktitle}
  {Progress in {Optics}}}},\ Vol.~\bibinfo {volume} {61},\ \bibinfo {editor}
  {edited by\ \bibinfo {editor} {\bibfnamefont {Taco~D.}\ \bibnamefont
  {Visser}}}\ (\bibinfo  {publisher} {Elsevier},\ \bibinfo {year} {2016})\ pp.\
  \bibinfo {pages} {113--236}\BibitemShut {NoStop}%
\bibitem [{\citenamefont {Gr{\" o}blacher}\ \emph {et~al.}(2009)\citenamefont
  {Gr{\" o}blacher}, \citenamefont {Hammerer}, \citenamefont {Vanner},\ and\
  \citenamefont {Aspelmeyer}}]{groblacher_observation_2009}%
  \BibitemOpen
  \bibfield  {author} {\bibinfo {author} {\bibfnamefont {Simon}\ \bibnamefont
  {Gr{\" o}blacher}}, \bibinfo {author} {\bibfnamefont {Klemens}\ \bibnamefont
  {Hammerer}}, \bibinfo {author} {\bibfnamefont {Michael~R.}\ \bibnamefont
  {Vanner}}, \ and\ \bibinfo {author} {\bibfnamefont {Markus}\ \bibnamefont
  {Aspelmeyer}},\ }\bibfield  {title} {\enquote {\bibinfo {title} {Observation
  of strong coupling between a micromechanical resonator and an optical cavity
  field},}\ }\href {\doibase 10.1038/nature08171} {\bibfield  {journal}
  {\bibinfo  {journal} {Nature}\ }\textbf {\bibinfo {volume} {460}},\ \bibinfo
  {pages} {724--727} (\bibinfo {year} {2009})}\BibitemShut {NoStop}%
\bibitem [{\citenamefont {Verhagen}\ \emph {et~al.}(2012)\citenamefont
  {Verhagen}, \citenamefont {Del{\' e}glise}, \citenamefont {Weis},
  \citenamefont {Schliesser},\ and\ \citenamefont
  {Kippenberg}}]{verhagen_quantum-coherent_2012}%
  \BibitemOpen
  \bibfield  {author} {\bibinfo {author} {\bibfnamefont {E.}~\bibnamefont
  {Verhagen}}, \bibinfo {author} {\bibfnamefont {S.}~\bibnamefont {Del{\'
  e}glise}}, \bibinfo {author} {\bibfnamefont {S.}~\bibnamefont {Weis}},
  \bibinfo {author} {\bibfnamefont {A.}~\bibnamefont {Schliesser}}, \ and\
  \bibinfo {author} {\bibfnamefont {T.~J.}\ \bibnamefont {Kippenberg}},\
  }\bibfield  {title} {\enquote {\bibinfo {title} {Quantum-coherent coupling of
  a mechanical oscillator to an optical cavity mode},}\ }\href {\doibase
  10.1038/nature10787} {\bibfield  {journal} {\bibinfo  {journal} {Nature}\
  }\textbf {\bibinfo {volume} {482}},\ \bibinfo {pages} {63--67} (\bibinfo
  {year} {2012})}\BibitemShut {NoStop}%
\bibitem [{\citenamefont {Weis}\ \emph {et~al.}(2010)\citenamefont {Weis},
  \citenamefont {Rivi{\` e}re}, \citenamefont {Del{\' e}glise}, \citenamefont
  {Gavartin}, \citenamefont {Arcizet}, \citenamefont {Schliesser},\ and\
  \citenamefont {Kippenberg}}]{weis_optomechanically_2010}%
  \BibitemOpen
  \bibfield  {author} {\bibinfo {author} {\bibfnamefont {Stefan}\ \bibnamefont
  {Weis}}, \bibinfo {author} {\bibfnamefont {R{\' e}mi}\ \bibnamefont {Rivi{\`
  e}re}}, \bibinfo {author} {\bibfnamefont {Samuel}\ \bibnamefont {Del{\'
  e}glise}}, \bibinfo {author} {\bibfnamefont {Emanuel}\ \bibnamefont
  {Gavartin}}, \bibinfo {author} {\bibfnamefont {Olivier}\ \bibnamefont
  {Arcizet}}, \bibinfo {author} {\bibfnamefont {Albert}\ \bibnamefont
  {Schliesser}}, \ and\ \bibinfo {author} {\bibfnamefont {Tobias~J.}\
  \bibnamefont {Kippenberg}},\ }\bibfield  {title} {\enquote {\bibinfo {title}
  {Optomechanically {Induced} {Transparency}},}\ }\href {\doibase
  10.1126/science.1195596} {\bibfield  {journal} {\bibinfo  {journal}
  {Science}\ }\textbf {\bibinfo {volume} {330}},\ \bibinfo {pages} {1520--1523}
  (\bibinfo {year} {2010})}\BibitemShut {NoStop}%
\bibitem [{\citenamefont {Chan}\ \emph {et~al.}(2011)\citenamefont {Chan},
  \citenamefont {Alegre}, \citenamefont {Safavi-Naeini}, \citenamefont {Hill},
  \citenamefont {Krause}, \citenamefont {Groeblacher}, \citenamefont
  {Aspelmeyer},\ and\ \citenamefont {Painter}}]{chan_laser_2011}%
  \BibitemOpen
  \bibfield  {author} {\bibinfo {author} {\bibfnamefont {Jasper}\ \bibnamefont
  {Chan}}, \bibinfo {author} {\bibfnamefont {T.~P.~Mayer}\ \bibnamefont
  {Alegre}}, \bibinfo {author} {\bibfnamefont {Amir~H.}\ \bibnamefont
  {Safavi-Naeini}}, \bibinfo {author} {\bibfnamefont {Jeff~T.}\ \bibnamefont
  {Hill}}, \bibinfo {author} {\bibfnamefont {Alex}\ \bibnamefont {Krause}},
  \bibinfo {author} {\bibfnamefont {Simon}\ \bibnamefont {Groeblacher}},
  \bibinfo {author} {\bibfnamefont {Markus}\ \bibnamefont {Aspelmeyer}}, \ and\
  \bibinfo {author} {\bibfnamefont {Oskar}\ \bibnamefont {Painter}},\
  }\bibfield  {title} {\enquote {\bibinfo {title} {Laser cooling of a
  nanomechanical oscillator into its quantum ground state},}\ }\href {\doibase
  10.1038/nature10461} {\bibfield  {journal} {\bibinfo  {journal} {Nature}\
  }\textbf {\bibinfo {volume} {478}},\ \bibinfo {pages} {89--92} (\bibinfo
  {year} {2011})},\ \bibinfo {note} {arXiv:1106.3614 [quant-ph]}\BibitemShut
  {NoStop}%
\bibitem [{\citenamefont {Teufel}\ \emph {et~al.}(2011)\citenamefont {Teufel},
  \citenamefont {Donner}, \citenamefont {Li}, \citenamefont {Harlow},
  \citenamefont {Allman}, \citenamefont {Cicak}, \citenamefont {Sirois},
  \citenamefont {Whittaker}, \citenamefont {Lehnert},\ and\ \citenamefont
  {Simmonds}}]{teufel_sideband_2011}%
  \BibitemOpen
  \bibfield  {author} {\bibinfo {author} {\bibfnamefont {J.~D.}\ \bibnamefont
  {Teufel}}, \bibinfo {author} {\bibfnamefont {T.}~\bibnamefont {Donner}},
  \bibinfo {author} {\bibfnamefont {Dale}\ \bibnamefont {Li}}, \bibinfo
  {author} {\bibfnamefont {J.~W.}\ \bibnamefont {Harlow}}, \bibinfo {author}
  {\bibfnamefont {M.~S.}\ \bibnamefont {Allman}}, \bibinfo {author}
  {\bibfnamefont {K.}~\bibnamefont {Cicak}}, \bibinfo {author} {\bibfnamefont
  {A.~J.}\ \bibnamefont {Sirois}}, \bibinfo {author} {\bibfnamefont {J.~D.}\
  \bibnamefont {Whittaker}}, \bibinfo {author} {\bibfnamefont {K.~W.}\
  \bibnamefont {Lehnert}}, \ and\ \bibinfo {author} {\bibfnamefont {R.~W.}\
  \bibnamefont {Simmonds}},\ }\bibfield  {title} {\enquote {\bibinfo {title}
  {Sideband cooling of micromechanical motion to the quantum ground state},}\
  }\href {\doibase 10.1038/nature10261} {\bibfield  {journal} {\bibinfo
  {journal} {Nature}\ }\textbf {\bibinfo {volume} {475}},\ \bibinfo {pages}
  {359--363} (\bibinfo {year} {2011})}\BibitemShut {NoStop}%
\bibitem [{\citenamefont {Wollman}\ \emph {et~al.}(2015)\citenamefont
  {Wollman}, \citenamefont {Lei}, \citenamefont {Weinstein}, \citenamefont
  {Suh}, \citenamefont {Kronwald}, \citenamefont {Marquardt}, \citenamefont
  {Clerk},\ and\ \citenamefont {Schwab}}]{wollman_quantum_2015}%
  \BibitemOpen
  \bibfield  {author} {\bibinfo {author} {\bibfnamefont {E.~E.}\ \bibnamefont
  {Wollman}}, \bibinfo {author} {\bibfnamefont {C.~U.}\ \bibnamefont {Lei}},
  \bibinfo {author} {\bibfnamefont {A.~J.}\ \bibnamefont {Weinstein}}, \bibinfo
  {author} {\bibfnamefont {J.}~\bibnamefont {Suh}}, \bibinfo {author}
  {\bibfnamefont {A.}~\bibnamefont {Kronwald}}, \bibinfo {author}
  {\bibfnamefont {F.}~\bibnamefont {Marquardt}}, \bibinfo {author}
  {\bibfnamefont {A.~A.}\ \bibnamefont {Clerk}}, \ and\ \bibinfo {author}
  {\bibfnamefont {K.~C.}\ \bibnamefont {Schwab}},\ }\bibfield  {title}
  {\enquote {\bibinfo {title} {Quantum squeezing of motion in a mechanical
  resonator},}\ }\href {\doibase 10.1126/science.aac5138} {\bibfield  {journal}
  {\bibinfo  {journal} {Science}\ }\textbf {\bibinfo {volume} {349}},\ \bibinfo
  {pages} {952--955} (\bibinfo {year} {2015})},\ \bibinfo {note} {arXiv:
  1507.01662}\BibitemShut {NoStop}%
\bibitem [{\citenamefont {Pirkkalainen}\ \emph {et~al.}(2015)\citenamefont
  {Pirkkalainen}, \citenamefont {Damsk{\" a}gg}, \citenamefont {Brandt},
  \citenamefont {Massel},\ and\ \citenamefont {Sillanp{\" a}{\"
  a}}}]{pirkkalainen_squeezing_2015}%
  \BibitemOpen
  \bibfield  {author} {\bibinfo {author} {\bibfnamefont {J.-M.}\ \bibnamefont
  {Pirkkalainen}}, \bibinfo {author} {\bibfnamefont {E.}~\bibnamefont {Damsk{\"
  a}gg}}, \bibinfo {author} {\bibfnamefont {M.}~\bibnamefont {Brandt}},
  \bibinfo {author} {\bibfnamefont {F.}~\bibnamefont {Massel}}, \ and\ \bibinfo
  {author} {\bibfnamefont {M.~A.}\ \bibnamefont {Sillanp{\" a}{\" a}}},\
  }\bibfield  {title} {\enquote {\bibinfo {title} {Squeezing of {Quantum}
  {Noise} of {Motion} in a {Micromechanical} {Resonator}},}\ }\href {\doibase
  10.1103/PhysRevLett.115.243601} {\bibfield  {journal} {\bibinfo  {journal}
  {Physical Review Letters}\ }\textbf {\bibinfo {volume} {115}},\ \bibinfo
  {pages} {243601} (\bibinfo {year} {2015})},\ \bibinfo {note} {arXiv:
  1507.04209}\BibitemShut {NoStop}%
\bibitem [{\citenamefont {Riedinger}\ \emph {et~al.}(2016)\citenamefont
  {Riedinger}, \citenamefont {Hong}, \citenamefont {Norte}, \citenamefont
  {Slater}, \citenamefont {Shang}, \citenamefont {Krause}, \citenamefont
  {Anant}, \citenamefont {Aspelmeyer},\ and\ \citenamefont {Gr{\"
  o}blacher}}]{riedinger_non-classical_2016}%
  \BibitemOpen
  \bibfield  {author} {\bibinfo {author} {\bibfnamefont {Ralf}\ \bibnamefont
  {Riedinger}}, \bibinfo {author} {\bibfnamefont {Sungkun}\ \bibnamefont
  {Hong}}, \bibinfo {author} {\bibfnamefont {Richard~A.}\ \bibnamefont
  {Norte}}, \bibinfo {author} {\bibfnamefont {Joshua~A.}\ \bibnamefont
  {Slater}}, \bibinfo {author} {\bibfnamefont {Juying}\ \bibnamefont {Shang}},
  \bibinfo {author} {\bibfnamefont {Alexander~G.}\ \bibnamefont {Krause}},
  \bibinfo {author} {\bibfnamefont {Vikas}\ \bibnamefont {Anant}}, \bibinfo
  {author} {\bibfnamefont {Markus}\ \bibnamefont {Aspelmeyer}}, \ and\ \bibinfo
  {author} {\bibfnamefont {Simon}\ \bibnamefont {Gr{\" o}blacher}},\ }\bibfield
   {title} {\enquote {\bibinfo {title} {Non-classical correlations between
  single photons and phonons from a mechanical oscillator},}\ }\href {\doibase
  10.1038/nature16536} {\bibfield  {journal} {\bibinfo  {journal} {Nature}\
  }\textbf {\bibinfo {volume} {530}},\ \bibinfo {pages} {313--316} (\bibinfo
  {year} {2016})},\ \bibinfo {note} {arXiv: 1512.05360}\BibitemShut {NoStop}%
\bibitem [{\citenamefont {Hong}\ \emph {et~al.}(2017)\citenamefont {Hong},
  \citenamefont {Riedinger}, \citenamefont {Marinkovi{\' c}}, \citenamefont
  {Wallucks}, \citenamefont {Hofer}, \citenamefont {Norte}, \citenamefont
  {Aspelmeyer},\ and\ \citenamefont {Gr{\" o}blacher}}]{hong_hanbury_2017}%
  \BibitemOpen
  \bibfield  {author} {\bibinfo {author} {\bibfnamefont {Sungkun}\ \bibnamefont
  {Hong}}, \bibinfo {author} {\bibfnamefont {Ralf}\ \bibnamefont {Riedinger}},
  \bibinfo {author} {\bibfnamefont {Igor}\ \bibnamefont {Marinkovi{\' c}}},
  \bibinfo {author} {\bibfnamefont {Andreas}\ \bibnamefont {Wallucks}},
  \bibinfo {author} {\bibfnamefont {Sebastian~G.}\ \bibnamefont {Hofer}},
  \bibinfo {author} {\bibfnamefont {Richard~A.}\ \bibnamefont {Norte}},
  \bibinfo {author} {\bibfnamefont {Markus}\ \bibnamefont {Aspelmeyer}}, \ and\
  \bibinfo {author} {\bibfnamefont {Simon}\ \bibnamefont {Gr{\" o}blacher}},\
  }\bibfield  {title} {\enquote {\bibinfo {title} {Hanbury {Brown} and {Twiss}
  interferometry of single phonons from an optomechanical resonator},}\ }\href
  {\doibase 10.1126/science.aan7939} {\bibfield  {journal} {\bibinfo  {journal}
  {Science}\ }\textbf {\bibinfo {volume} {358}},\ \bibinfo {pages} {203--206}
  (\bibinfo {year} {2017})},\ \bibinfo {note} {arXiv: 1706.03777}\BibitemShut
  {NoStop}%
\bibitem [{\citenamefont {Palomaki}\ \emph {et~al.}(2013)\citenamefont
  {Palomaki}, \citenamefont {Teufel}, \citenamefont {Simmonds},\ and\
  \citenamefont {Lehnert}}]{palomaki_entangling_2013}%
  \BibitemOpen
  \bibfield  {author} {\bibinfo {author} {\bibfnamefont {T.~A.}\ \bibnamefont
  {Palomaki}}, \bibinfo {author} {\bibfnamefont {J.~D.}\ \bibnamefont
  {Teufel}}, \bibinfo {author} {\bibfnamefont {R.~W.}\ \bibnamefont
  {Simmonds}}, \ and\ \bibinfo {author} {\bibfnamefont {K.~W.}\ \bibnamefont
  {Lehnert}},\ }\bibfield  {title} {\enquote {\bibinfo {title} {Entangling
  {Mechanical} {Motion} with {Microwave} {Fields}},}\ }\href {\doibase
  10.1126/science.1244563} {\bibfield  {journal} {\bibinfo  {journal}
  {Science}\ }\textbf {\bibinfo {volume} {342}},\ \bibinfo {pages} {710--713}
  (\bibinfo {year} {2013})}\BibitemShut {NoStop}%
\bibitem [{\citenamefont {Riedinger}\ \emph {et~al.}(2018)\citenamefont
  {Riedinger}, \citenamefont {Wallucks}, \citenamefont {Marinkovi{\' c}},
  \citenamefont {L{\" o}schnauer}, \citenamefont {Aspelmeyer}, \citenamefont
  {Hong},\ and\ \citenamefont {Gr{\" o}blacher}}]{riedinger_remote_2018}%
  \BibitemOpen
  \bibfield  {author} {\bibinfo {author} {\bibfnamefont {Ralf}\ \bibnamefont
  {Riedinger}}, \bibinfo {author} {\bibfnamefont {Andreas}\ \bibnamefont
  {Wallucks}}, \bibinfo {author} {\bibfnamefont {Igor}\ \bibnamefont
  {Marinkovi{\' c}}}, \bibinfo {author} {\bibfnamefont {Clemens}\ \bibnamefont
  {L{\" o}schnauer}}, \bibinfo {author} {\bibfnamefont {Markus}\ \bibnamefont
  {Aspelmeyer}}, \bibinfo {author} {\bibfnamefont {Sungkun}\ \bibnamefont
  {Hong}}, \ and\ \bibinfo {author} {\bibfnamefont {Simon}\ \bibnamefont {Gr{\"
  o}blacher}},\ }\bibfield  {title} {\enquote {\bibinfo {title} {Remote quantum
  entanglement between two micromechanical oscillators},}\ }\href {\doibase
  10.1038/s41586-018-0036-z} {\bibfield  {journal} {\bibinfo  {journal}
  {Nature}\ }\textbf {\bibinfo {volume} {556}},\ \bibinfo {pages} {473--477}
  (\bibinfo {year} {2018})},\ \bibinfo {note} {arXiv: 1710.11147}\BibitemShut
  {NoStop}%
\bibitem [{\citenamefont {Ockeloen-Korppi}\ \emph {et~al.}(2018)\citenamefont
  {Ockeloen-Korppi}, \citenamefont {Damsk{\" a}gg}, \citenamefont
  {Pirkkalainen}, \citenamefont {Asjad}, \citenamefont {Clerk}, \citenamefont
  {Massel}, \citenamefont {Woolley},\ and\ \citenamefont {Sillanp{\" a}{\"
  a}}}]{ockeloen-korppi_stabilized_2018}%
  \BibitemOpen
  \bibfield  {author} {\bibinfo {author} {\bibfnamefont {C.~F.}\ \bibnamefont
  {Ockeloen-Korppi}}, \bibinfo {author} {\bibfnamefont {E.}~\bibnamefont
  {Damsk{\" a}gg}}, \bibinfo {author} {\bibfnamefont {J.-M.}\ \bibnamefont
  {Pirkkalainen}}, \bibinfo {author} {\bibfnamefont {M.}~\bibnamefont {Asjad}},
  \bibinfo {author} {\bibfnamefont {A.~A.}\ \bibnamefont {Clerk}}, \bibinfo
  {author} {\bibfnamefont {F.}~\bibnamefont {Massel}}, \bibinfo {author}
  {\bibfnamefont {M.~J.}\ \bibnamefont {Woolley}}, \ and\ \bibinfo {author}
  {\bibfnamefont {M.~A.}\ \bibnamefont {Sillanp{\" a}{\" a}}},\ }\bibfield
  {title} {\enquote {\bibinfo {title} {Stabilized entanglement of massive
  mechanical oscillators},}\ }\href {\doibase 10.1038/s41586-018-0038-x}
  {\bibfield  {journal} {\bibinfo  {journal} {Nature}\ }\textbf {\bibinfo
  {volume} {556}},\ \bibinfo {pages} {478--482} (\bibinfo {year} {2018})},\
  \bibinfo {note} {arXiv: 1711.01640}\BibitemShut {NoStop}%
\bibitem [{\citenamefont {Hofer}\ \emph {et~al.}(2011)\citenamefont {Hofer},
  \citenamefont {Wieczorek}, \citenamefont {Aspelmeyer},\ and\ \citenamefont
  {Hammerer}}]{hofer_quantum_2011}%
  \BibitemOpen
  \bibfield  {author} {\bibinfo {author} {\bibfnamefont {Sebastian~G.}\
  \bibnamefont {Hofer}}, \bibinfo {author} {\bibfnamefont {Witlef}\
  \bibnamefont {Wieczorek}}, \bibinfo {author} {\bibfnamefont {Markus}\
  \bibnamefont {Aspelmeyer}}, \ and\ \bibinfo {author} {\bibfnamefont
  {Klemens}\ \bibnamefont {Hammerer}},\ }\bibfield  {title} {\enquote {\bibinfo
  {title} {Quantum entanglement and teleportation in pulsed cavity
  optomechanics},}\ }\href {\doibase 10.1103/PhysRevA.84.052327} {\bibfield
  {journal} {\bibinfo  {journal} {Physical Review A}\ }\textbf {\bibinfo
  {volume} {84}},\ \bibinfo {pages} {052327} (\bibinfo {year} {2011})},\
  \bibinfo {note} {arXiv: 1108.2586}\BibitemShut {NoStop}%
\bibitem [{\citenamefont {Vanner}\ \emph {et~al.}(2011)\citenamefont {Vanner},
  \citenamefont {Pikovski}, \citenamefont {Cole}, \citenamefont {Kim},
  \citenamefont {Brukner}, \citenamefont {Hammerer}, \citenamefont {Milburn},\
  and\ \citenamefont {Aspelmeyer}}]{vanner_pulsed_2011}%
  \BibitemOpen
  \bibfield  {author} {\bibinfo {author} {\bibfnamefont {M.~R.}\ \bibnamefont
  {Vanner}}, \bibinfo {author} {\bibfnamefont {I.}~\bibnamefont {Pikovski}},
  \bibinfo {author} {\bibfnamefont {G.~D.}\ \bibnamefont {Cole}}, \bibinfo
  {author} {\bibfnamefont {M.~S.}\ \bibnamefont {Kim}}, \bibinfo {author}
  {\bibfnamefont {{\v C}}~\bibnamefont {Brukner}}, \bibinfo {author}
  {\bibfnamefont {K.}~\bibnamefont {Hammerer}}, \bibinfo {author}
  {\bibfnamefont {G.~J.}\ \bibnamefont {Milburn}}, \ and\ \bibinfo {author}
  {\bibfnamefont {M.}~\bibnamefont {Aspelmeyer}},\ }\bibfield  {title}
  {\enquote {\bibinfo {title} {Pulsed quantum optomechanics},}\ }\href
  {\doibase 10.1073/pnas.1105098108} {\bibfield  {journal} {\bibinfo  {journal}
  {Proceedings of the National Academy of Sciences}\ }\textbf {\bibinfo
  {volume} {108}},\ \bibinfo {pages} {16182--16187} (\bibinfo {year} {2011})},\
  \bibinfo {note} {arXiv:1011.0879 [cond-mat, physics:quant-ph]}\BibitemShut
  {NoStop}%
\bibitem [{\citenamefont {Braginsky}\ and\ \citenamefont
  {Minakova}(1964)}]{braginsky_notitle_1964}%
  \BibitemOpen
  \bibfield  {author} {\bibinfo {author} {\bibfnamefont {V.~B.}\ \bibnamefont
  {Braginsky}}\ and\ \bibinfo {author} {\bibfnamefont {I.~I.}\ \bibnamefont
  {Minakova}},\ }\href@noop {} {\bibfield  {journal} {\bibinfo  {journal}
  {Vestnik Moskovskogo Universiteta, Seriya 3}\ }\textbf {\bibinfo {volume}
  {1}},\ \bibinfo {pages} {69} (\bibinfo {year} {1964})},\ \bibinfo {note} {(in
  Russian)}\BibitemShut {NoStop}%
\bibitem [{\citenamefont {Braginsky}\ and\ \citenamefont
  {Manukin}(1967)}]{braginsky_ponderomotive_1967}%
  \BibitemOpen
  \bibfield  {author} {\bibinfo {author} {\bibfnamefont {Vladimir~Borisovich}\
  \bibnamefont {Braginsky}}\ and\ \bibinfo {author} {\bibfnamefont {A.~B.}\
  \bibnamefont {Manukin}},\ }\bibfield  {title} {\enquote {\bibinfo {title}
  {Ponderomotive {Effects} of {Electromagnetic} {Radiation}},}\ }\href@noop {}
  {\bibfield  {journal} {\bibinfo  {journal} {JETP}\ }\textbf {\bibinfo
  {volume} {25}},\ \bibinfo {pages} {653} (\bibinfo {year} {1967})}\BibitemShut
  {NoStop}%
\bibitem [{\citenamefont {Rugar}\ \emph {et~al.}(2004)\citenamefont {Rugar},
  \citenamefont {Budakian}, \citenamefont {Mamin},\ and\ \citenamefont
  {Chui}}]{rugar_single_2004}%
  \BibitemOpen
  \bibfield  {author} {\bibinfo {author} {\bibfnamefont {D.}~\bibnamefont
  {Rugar}}, \bibinfo {author} {\bibfnamefont {R.}~\bibnamefont {Budakian}},
  \bibinfo {author} {\bibfnamefont {H.~J.}\ \bibnamefont {Mamin}}, \ and\
  \bibinfo {author} {\bibfnamefont {B.~W.}\ \bibnamefont {Chui}},\ }\bibfield
  {title} {\enquote {\bibinfo {title} {Single spin detection by magnetic
  resonance force microscopy},}\ }\href {\doibase 10.1038/nature02658}
  {\bibfield  {journal} {\bibinfo  {journal} {Nature}\ }\textbf {\bibinfo
  {volume} {430}},\ \bibinfo {pages} {329--332} (\bibinfo {year}
  {2004})}\BibitemShut {NoStop}%
\bibitem [{\citenamefont {Gavartin}\ \emph {et~al.}(2012)\citenamefont
  {Gavartin}, \citenamefont {Verlot},\ and\ \citenamefont
  {Kippenberg}}]{gavartin_hybrid_2012}%
  \BibitemOpen
  \bibfield  {author} {\bibinfo {author} {\bibfnamefont {E.}~\bibnamefont
  {Gavartin}}, \bibinfo {author} {\bibfnamefont {P.}~\bibnamefont {Verlot}}, \
  and\ \bibinfo {author} {\bibfnamefont {T.~J.}\ \bibnamefont {Kippenberg}},\
  }\bibfield  {title} {\enquote {\bibinfo {title} {A hybrid on-chip
  optomechanical transducer for ultrasensitive force measurements},}\ }\href
  {\doibase 10.1038/nnano.2012.97} {\bibfield  {journal} {\bibinfo  {journal}
  {Nature Nanotechnology}\ }\textbf {\bibinfo {volume} {7}},\ \bibinfo {pages}
  {509--514} (\bibinfo {year} {2012})}\BibitemShut {NoStop}%
\bibitem [{\citenamefont {Seveso}\ \emph {et~al.}(2017)\citenamefont {Seveso},
  \citenamefont {Peri},\ and\ \citenamefont {Paris}}]{seveso_quantum_2017}%
  \BibitemOpen
  \bibfield  {author} {\bibinfo {author} {\bibfnamefont {Luigi}\ \bibnamefont
  {Seveso}}, \bibinfo {author} {\bibfnamefont {Valerio}\ \bibnamefont {Peri}},
  \ and\ \bibinfo {author} {\bibfnamefont {Matteo G.~A.}\ \bibnamefont
  {Paris}},\ }\bibfield  {title} {\enquote {\bibinfo {title} {Quantum limits to
  mass sensing in a gravitational field},}\ }\href {\doibase
  10.1088/1751-8121/aa6cc5} {\bibfield  {journal} {\bibinfo  {journal} {Journal
  of Physics A: Mathematical and Theoretical}\ }\textbf {\bibinfo {volume}
  {50}},\ \bibinfo {pages} {235301} (\bibinfo {year} {2017})}\BibitemShut
  {NoStop}%
\bibitem [{\citenamefont {Abbott}\ \emph {et~al.}(2009)\citenamefont {Abbott}
  \emph {et~al.}}]{Abbott:2007kv}%
  \BibitemOpen
  \bibfield  {author} {\bibinfo {author} {\bibfnamefont {B.~P.}\ \bibnamefont
  {Abbott}} \emph {et~al.},\ }\bibfield  {title} {\enquote {\bibinfo {title}
  {{LIGO: The Laser interferometer gravitational-wave observatory}},}\ }\href
  {\doibase 10.1088/0034-4885/72/7/076901} {\bibfield  {journal} {\bibinfo
  {journal} {Rept. Prog. Phys.}\ }\textbf {\bibinfo {volume} {72}},\ \bibinfo
  {pages} {076901} (\bibinfo {year} {2009})},\ \Eprint
  {http://arxiv.org/abs/0711.3041} {0711.3041} \BibitemShut {NoStop}%
\bibitem [{\citenamefont {{LIGO Scientific Collaboration and Virgo
  Collaboration}}(2016)}]{ligo_scientific_collaboration_and_virgo_collaboration_observation_2016}%
  \BibitemOpen
  \bibfield  {author} {\bibinfo {author} {\bibnamefont {{LIGO Scientific
  Collaboration and Virgo Collaboration}}},\ }\bibfield  {title} {\enquote
  {\bibinfo {title} {Observation of {Gravitational} {Waves} from a {Binary}
  {Black} {Hole} {Merger}},}\ }\href {\doibase 10.1103/PhysRevLett.116.061102}
  {\bibfield  {journal} {\bibinfo  {journal} {Physical Review Letters}\
  }\textbf {\bibinfo {volume} {116}},\ \bibinfo {pages} {061102} (\bibinfo
  {year} {2016})},\ \bibinfo {note} {arXiv: 1602.03837}\BibitemShut {NoStop}%
\bibitem [{\citenamefont {Forstner}\ \emph {et~al.}(2012)\citenamefont
  {Forstner}, \citenamefont {Prams}, \citenamefont {Knittel}, \citenamefont
  {van Ooijen}, \citenamefont {Swaim}, \citenamefont {Harris}, \citenamefont
  {Szorkovszky}, \citenamefont {Bowen},\ and\ \citenamefont
  {Rubinsztein-Dunlop}}]{forstner_cavity_2012}%
  \BibitemOpen
  \bibfield  {author} {\bibinfo {author} {\bibfnamefont {S.}~\bibnamefont
  {Forstner}}, \bibinfo {author} {\bibfnamefont {S.}~\bibnamefont {Prams}},
  \bibinfo {author} {\bibfnamefont {J.}~\bibnamefont {Knittel}}, \bibinfo
  {author} {\bibfnamefont {E.~D.}\ \bibnamefont {van Ooijen}}, \bibinfo
  {author} {\bibfnamefont {J.~D.}\ \bibnamefont {Swaim}}, \bibinfo {author}
  {\bibfnamefont {G.~I.}\ \bibnamefont {Harris}}, \bibinfo {author}
  {\bibfnamefont {A.}~\bibnamefont {Szorkovszky}}, \bibinfo {author}
  {\bibfnamefont {W.~P.}\ \bibnamefont {Bowen}}, \ and\ \bibinfo {author}
  {\bibfnamefont {H.}~\bibnamefont {Rubinsztein-Dunlop}},\ }\bibfield  {title}
  {\enquote {\bibinfo {title} {Cavity {Optomechanical} {Magnetometer}},}\
  }\href {\doibase 10.1103/PhysRevLett.108.120801} {\bibfield  {journal}
  {\bibinfo  {journal} {Physical Review Letters}\ }\textbf {\bibinfo {volume}
  {108}},\ \bibinfo {pages} {120801} (\bibinfo {year} {2012})}\BibitemShut
  {NoStop}%
\bibitem [{\citenamefont {Yu}\ \emph {et~al.}(2016)\citenamefont {Yu},
  \citenamefont {Janousek}, \citenamefont {Sheridan}, \citenamefont {McAuslan},
  \citenamefont {Rubinsztein-Dunlop}, \citenamefont {Lam}, \citenamefont
  {Zhang},\ and\ \citenamefont {Bowen}}]{yu_optomechanical_2016}%
  \BibitemOpen
  \bibfield  {author} {\bibinfo {author} {\bibfnamefont {Changqiu}\
  \bibnamefont {Yu}}, \bibinfo {author} {\bibfnamefont {Jiri}\ \bibnamefont
  {Janousek}}, \bibinfo {author} {\bibfnamefont {Eoin}\ \bibnamefont
  {Sheridan}}, \bibinfo {author} {\bibfnamefont {David~L.}\ \bibnamefont
  {McAuslan}}, \bibinfo {author} {\bibfnamefont {Halina}\ \bibnamefont
  {Rubinsztein-Dunlop}}, \bibinfo {author} {\bibfnamefont {Ping~Koy}\
  \bibnamefont {Lam}}, \bibinfo {author} {\bibfnamefont {Yundong}\ \bibnamefont
  {Zhang}}, \ and\ \bibinfo {author} {\bibfnamefont {Warwick~P.}\ \bibnamefont
  {Bowen}},\ }\bibfield  {title} {\enquote {\bibinfo {title} {Optomechanical
  {Magnetometry} with a {Macroscopic} {Resonator}},}\ }\href {\doibase
  10.1103/PhysRevApplied.5.044007} {\bibfield  {journal} {\bibinfo  {journal}
  {Physical Review Applied}\ }\textbf {\bibinfo {volume} {5}},\ \bibinfo
  {pages} {044007} (\bibinfo {year} {2016})}\BibitemShut {NoStop}%
\bibitem [{\citenamefont {Li}\ \emph {et~al.}(2018)\citenamefont {Li},
  \citenamefont {Bilek}, \citenamefont {Hoff}, \citenamefont {Madsen},
  \citenamefont {Forstner}, \citenamefont {Prakash}, \citenamefont {Sch{\"
  a}fermeier}, \citenamefont {Gehring}, \citenamefont {Bowen},\ and\
  \citenamefont {Andersen}}]{li_quantum_2018}%
  \BibitemOpen
  \bibfield  {author} {\bibinfo {author} {\bibfnamefont {Bei-Bei}\ \bibnamefont
  {Li}}, \bibinfo {author} {\bibfnamefont {Jan}\ \bibnamefont {Bilek}},
  \bibinfo {author} {\bibfnamefont {Ulrich~B.}\ \bibnamefont {Hoff}}, \bibinfo
  {author} {\bibfnamefont {Lars~S.}\ \bibnamefont {Madsen}}, \bibinfo {author}
  {\bibfnamefont {Stefan}\ \bibnamefont {Forstner}}, \bibinfo {author}
  {\bibfnamefont {Varun}\ \bibnamefont {Prakash}}, \bibinfo {author}
  {\bibfnamefont {Clemens}\ \bibnamefont {Sch{\" a}fermeier}}, \bibinfo
  {author} {\bibfnamefont {Tobias}\ \bibnamefont {Gehring}}, \bibinfo {author}
  {\bibfnamefont {Warwick~P.}\ \bibnamefont {Bowen}}, \ and\ \bibinfo {author}
  {\bibfnamefont {Ulrik~L.}\ \bibnamefont {Andersen}},\ }\bibfield  {title}
  {\enquote {\bibinfo {title} {Quantum enhanced optomechanical magnetometry},}\
  }\href@noop {} {\bibfield  {journal} {\bibinfo  {journal} {arXiv:1802.09738
  [physics, physics:quant-ph]}\ } (\bibinfo {year} {2018})},\ \bibinfo {note}
  {arXiv: 1802.09738}\BibitemShut {NoStop}%
\bibitem [{\citenamefont {Arndt}\ \emph {et~al.}(2009)\citenamefont {Arndt},
  \citenamefont {Juffmann},\ and\ \citenamefont {Vedral}}]{arndt_quantum_2009}%
  \BibitemOpen
  \bibfield  {author} {\bibinfo {author} {\bibfnamefont {Markus}\ \bibnamefont
  {Arndt}}, \bibinfo {author} {\bibfnamefont {Thomas}\ \bibnamefont
  {Juffmann}}, \ and\ \bibinfo {author} {\bibfnamefont {Vlatko}\ \bibnamefont
  {Vedral}},\ }\bibfield  {title} {\enquote {\bibinfo {title} {Quantum physics
  meets biology},}\ }\href {\doibase 10.2976/1.3244985} {\bibfield  {journal}
  {\bibinfo  {journal} {HFSP Journal}\ }\textbf {\bibinfo {volume} {3}},\
  \bibinfo {pages} {386--400} (\bibinfo {year} {2009})}\BibitemShut {NoStop}%
\bibitem [{\citenamefont {Bagci}\ \emph {et~al.}(2014)\citenamefont {Bagci},
  \citenamefont {Simonsen}, \citenamefont {Schmid}, \citenamefont {Villanueva},
  \citenamefont {Zeuthen}, \citenamefont {Appel}, \citenamefont {Taylor},
  \citenamefont {S\o{}rensen}, \citenamefont {Usami}, \citenamefont
  {Schliesser},\ and\ \citenamefont {Polzik}}]{bagci_optical_2014}%
  \BibitemOpen
  \bibfield  {author} {\bibinfo {author} {\bibfnamefont {T.}~\bibnamefont
  {Bagci}}, \bibinfo {author} {\bibfnamefont {A.}~\bibnamefont {Simonsen}},
  \bibinfo {author} {\bibfnamefont {S.}~\bibnamefont {Schmid}}, \bibinfo
  {author} {\bibfnamefont {L.~G.}\ \bibnamefont {Villanueva}}, \bibinfo
  {author} {\bibfnamefont {E.}~\bibnamefont {Zeuthen}}, \bibinfo {author}
  {\bibfnamefont {J.}~\bibnamefont {Appel}}, \bibinfo {author} {\bibfnamefont
  {J.~M.}\ \bibnamefont {Taylor}}, \bibinfo {author} {\bibfnamefont
  {A.}~\bibnamefont {S\o{}rensen}}, \bibinfo {author} {\bibfnamefont
  {K.}~\bibnamefont {Usami}}, \bibinfo {author} {\bibfnamefont
  {A.}~\bibnamefont {Schliesser}}, \ and\ \bibinfo {author} {\bibfnamefont
  {E.~S.}\ \bibnamefont {Polzik}},\ }\bibfield  {title} {\enquote {\bibinfo
  {title} {Optical detection of radio waves through a nanomechanical
  transducer},}\ }\href {\doibase 10.1038/nature13029} {\bibfield  {journal}
  {\bibinfo  {journal} {Nature}\ }\textbf {\bibinfo {volume} {507}},\ \bibinfo
  {pages} {81--85} (\bibinfo {year} {2014})}\BibitemShut {NoStop}%
\bibitem [{\citenamefont {Andrews}\ \emph {et~al.}(2014)\citenamefont
  {Andrews}, \citenamefont {Peterson}, \citenamefont {Purdy}, \citenamefont
  {Cicak}, \citenamefont {Simmonds}, \citenamefont {Regal},\ and\ \citenamefont
  {Lehnert}}]{andrews_bidirectional_2014}%
  \BibitemOpen
  \bibfield  {author} {\bibinfo {author} {\bibfnamefont {R.~W.}\ \bibnamefont
  {Andrews}}, \bibinfo {author} {\bibfnamefont {R.~W.}\ \bibnamefont
  {Peterson}}, \bibinfo {author} {\bibfnamefont {T.~P.}\ \bibnamefont {Purdy}},
  \bibinfo {author} {\bibfnamefont {K.}~\bibnamefont {Cicak}}, \bibinfo
  {author} {\bibfnamefont {R.~W.}\ \bibnamefont {Simmonds}}, \bibinfo {author}
  {\bibfnamefont {C.~A.}\ \bibnamefont {Regal}}, \ and\ \bibinfo {author}
  {\bibfnamefont {K.~W.}\ \bibnamefont {Lehnert}},\ }\bibfield  {title}
  {\enquote {\bibinfo {title} {Bidirectional and efficient conversion between
  microwave and optical light},}\ }\href {\doibase 10.1038/nphys2911}
  {\bibfield  {journal} {\bibinfo  {journal} {Nature Physics}\ }\textbf
  {\bibinfo {volume} {10}},\ \bibinfo {pages} {321--326} (\bibinfo {year}
  {2014})},\ \bibinfo {note} {arXiv: 1310.5276}\BibitemShut {NoStop}%
\bibitem [{\citenamefont {Andrews}\ \emph {et~al.}(2015)\citenamefont
  {Andrews}, \citenamefont {Reed}, \citenamefont {Cicak}, \citenamefont
  {Teufel},\ and\ \citenamefont {Lehnert}}]{andrews_quantum-enabled_2015}%
  \BibitemOpen
  \bibfield  {author} {\bibinfo {author} {\bibfnamefont {R.~W.}\ \bibnamefont
  {Andrews}}, \bibinfo {author} {\bibfnamefont {A.~P.}\ \bibnamefont {Reed}},
  \bibinfo {author} {\bibfnamefont {K.}~\bibnamefont {Cicak}}, \bibinfo
  {author} {\bibfnamefont {J.~D.}\ \bibnamefont {Teufel}}, \ and\ \bibinfo
  {author} {\bibfnamefont {K.~W.}\ \bibnamefont {Lehnert}},\ }\bibfield
  {title} {\enquote {\bibinfo {title} {Quantum-enabled temporal and spectral
  mode conversion of microwave signals},}\ }\href {\doibase
  10.1038/ncomms10021} {\bibfield  {journal} {\bibinfo  {journal} {Nature
  Communications}\ }\textbf {\bibinfo {volume} {6}},\ \bibinfo {pages} {10021}
  (\bibinfo {year} {2015})},\ \bibinfo {note} {arXiv: 1506.02296}\BibitemShut
  {NoStop}%
\bibitem [{\citenamefont {Lecocq}\ \emph {et~al.}(2016)\citenamefont {Lecocq},
  \citenamefont {Clark}, \citenamefont {Simmonds}, \citenamefont {Aumentado},\
  and\ \citenamefont {Teufel}}]{lecocq_mechanically_2016}%
  \BibitemOpen
  \bibfield  {author} {\bibinfo {author} {\bibfnamefont {F.}~\bibnamefont
  {Lecocq}}, \bibinfo {author} {\bibfnamefont {J.~B.}\ \bibnamefont {Clark}},
  \bibinfo {author} {\bibfnamefont {R.~W.}\ \bibnamefont {Simmonds}}, \bibinfo
  {author} {\bibfnamefont {J.}~\bibnamefont {Aumentado}}, \ and\ \bibinfo
  {author} {\bibfnamefont {J.~D.}\ \bibnamefont {Teufel}},\ }\bibfield  {title}
  {\enquote {\bibinfo {title} {Mechanically {Mediated} {Microwave} {Frequency}
  {Conversion} in the {Quantum} {Regime}},}\ }\href {\doibase
  10.1103/PhysRevLett.116.043601} {\bibfield  {journal} {\bibinfo  {journal}
  {Physical Review Letters}\ }\textbf {\bibinfo {volume} {116}},\ \bibinfo
  {pages} {043601} (\bibinfo {year} {2016})},\ \bibinfo {note} {arXiv:
  1512.00078}\BibitemShut {NoStop}%
\bibitem [{\citenamefont {Peterson}\ \emph {et~al.}(2017)\citenamefont
  {Peterson}, \citenamefont {Lecocq}, \citenamefont {Cicak}, \citenamefont
  {Simmonds}, \citenamefont {Aumentado},\ and\ \citenamefont
  {Teufel}}]{peterson_demonstration_2017}%
  \BibitemOpen
  \bibfield  {author} {\bibinfo {author} {\bibfnamefont {G.~A.}\ \bibnamefont
  {Peterson}}, \bibinfo {author} {\bibfnamefont {F.}~\bibnamefont {Lecocq}},
  \bibinfo {author} {\bibfnamefont {K.}~\bibnamefont {Cicak}}, \bibinfo
  {author} {\bibfnamefont {R.~W.}\ \bibnamefont {Simmonds}}, \bibinfo {author}
  {\bibfnamefont {J.}~\bibnamefont {Aumentado}}, \ and\ \bibinfo {author}
  {\bibfnamefont {J.~D.}\ \bibnamefont {Teufel}},\ }\bibfield  {title}
  {\enquote {\bibinfo {title} {Demonstration of {Efficient} {Nonreciprocity} in
  a {Microwave} {Optomechanical} {Circuit}},}\ }\href {\doibase
  10.1103/PhysRevX.7.031001} {\bibfield  {journal} {\bibinfo  {journal}
  {Physical Review X}\ }\textbf {\bibinfo {volume} {7}},\ \bibinfo {pages}
  {031001} (\bibinfo {year} {2017})},\ \bibinfo {note} {arXiv:
  1703.05269}\BibitemShut {NoStop}%
\bibitem [{\citenamefont {Barzanjeh}\ \emph {et~al.}(2017)\citenamefont
  {Barzanjeh}, \citenamefont {Wulf}, \citenamefont {Peruzzo}, \citenamefont
  {Kalaee}, \citenamefont {Dieterle}, \citenamefont {Painter},\ and\
  \citenamefont {Fink}}]{barzanjeh_mechanical_2017}%
  \BibitemOpen
  \bibfield  {author} {\bibinfo {author} {\bibfnamefont {S.}~\bibnamefont
  {Barzanjeh}}, \bibinfo {author} {\bibfnamefont {M.}~\bibnamefont {Wulf}},
  \bibinfo {author} {\bibfnamefont {M.}~\bibnamefont {Peruzzo}}, \bibinfo
  {author} {\bibfnamefont {M.}~\bibnamefont {Kalaee}}, \bibinfo {author}
  {\bibfnamefont {P.~B.}\ \bibnamefont {Dieterle}}, \bibinfo {author}
  {\bibfnamefont {O.}~\bibnamefont {Painter}}, \ and\ \bibinfo {author}
  {\bibfnamefont {J.~M.}\ \bibnamefont {Fink}},\ }\bibfield  {title} {\enquote
  {\bibinfo {title} {Mechanical on-chip microwave circulator},}\ }\href
  {\doibase 10.1038/s41467-017-01304-x} {\bibfield  {journal} {\bibinfo
  {journal} {Nature Communications}\ }\textbf {\bibinfo {volume} {8}},\
  \bibinfo {pages} {953} (\bibinfo {year} {2017})},\ \bibinfo {note} {arXiv:
  1706.00376}\BibitemShut {NoStop}%
\bibitem [{\citenamefont {Malz}\ \emph {et~al.}(2018)\citenamefont {Malz},
  \citenamefont {T{\' o}th}, \citenamefont {Bernier}, \citenamefont {Feofanov},
  \citenamefont {Kippenberg},\ and\ \citenamefont
  {Nunnenkamp}}]{malz_quantum-limited_2018}%
  \BibitemOpen
  \bibfield  {author} {\bibinfo {author} {\bibfnamefont {Daniel}\ \bibnamefont
  {Malz}}, \bibinfo {author} {\bibfnamefont {L{\' a}szl{\' o}~D.}\ \bibnamefont
  {T{\' o}th}}, \bibinfo {author} {\bibfnamefont {Nathan~R.}\ \bibnamefont
  {Bernier}}, \bibinfo {author} {\bibfnamefont {Alexey~K.}\ \bibnamefont
  {Feofanov}}, \bibinfo {author} {\bibfnamefont {Tobias~J.}\ \bibnamefont
  {Kippenberg}}, \ and\ \bibinfo {author} {\bibfnamefont {Andreas}\
  \bibnamefont {Nunnenkamp}},\ }\bibfield  {title} {\enquote {\bibinfo {title}
  {Quantum-{Limited} {Directional} {Amplifiers} with {Optomechanics}},}\ }\href
  {\doibase 10.1103/PhysRevLett.120.023601} {\bibfield  {journal} {\bibinfo
  {journal} {Physical Review Letters}\ }\textbf {\bibinfo {volume} {120}},\
  \bibinfo {pages} {023601} (\bibinfo {year} {2018})},\ \bibinfo {note} {arXiv:
  1705.00436}\BibitemShut {NoStop}%
\bibitem [{\citenamefont {Ruesink}\ \emph {et~al.}(2018)\citenamefont
  {Ruesink}, \citenamefont {Mathew}, \citenamefont {Miri}, \citenamefont {Al{\`
  u}},\ and\ \citenamefont {Verhagen}}]{ruesink_optical_2018}%
  \BibitemOpen
  \bibfield  {author} {\bibinfo {author} {\bibfnamefont {Freek}\ \bibnamefont
  {Ruesink}}, \bibinfo {author} {\bibfnamefont {John~P.}\ \bibnamefont
  {Mathew}}, \bibinfo {author} {\bibfnamefont {Mohammad-Ali}\ \bibnamefont
  {Miri}}, \bibinfo {author} {\bibfnamefont {Andrea}\ \bibnamefont {Al{\` u}}},
  \ and\ \bibinfo {author} {\bibfnamefont {Ewold}\ \bibnamefont {Verhagen}},\
  }\bibfield  {title} {\enquote {\bibinfo {title} {Optical circulation in a
  multimode optomechanical resonator},}\ }\href {\doibase
  10.1038/s41467-018-04202-y} {\bibfield  {journal} {\bibinfo  {journal}
  {Nature Communications}\ }\textbf {\bibinfo {volume} {9}},\ \bibinfo {pages}
  {1798} (\bibinfo {year} {2018})},\ \bibinfo {note} {arXiv:
  1708.07792}\BibitemShut {NoStop}%
\bibitem [{\citenamefont {Kaltenbaek}\ \emph {et~al.}(2016)\citenamefont
  {Kaltenbaek}, \citenamefont {Aspelmeyer}, \citenamefont {Barker},
  \citenamefont {Bassi}, \citenamefont {Bateman}, \citenamefont {Bongs},
  \citenamefont {Bose}, \citenamefont {Braxmaier}, \citenamefont {Brukner},
  \citenamefont {Christophe}, \citenamefont {Chwalla}, \citenamefont {Cohadon},
  \citenamefont {Cruise}, \citenamefont {Curceanu}, \citenamefont {Dholakia},
  \citenamefont {Di{\' o}si}, \citenamefont {D{\" o}ringshoff}, \citenamefont
  {Ertmer}, \citenamefont {Gieseler}, \citenamefont {G{\" u}rlebeck},
  \citenamefont {Hechenblaikner}, \citenamefont {Heidmann}, \citenamefont
  {Herrmann}, \citenamefont {Hossenfelder}, \citenamefont {Johann},
  \citenamefont {Kiesel}, \citenamefont {Kim}, \citenamefont {L{\" a}mmerzahl},
  \citenamefont {Lambrecht}, \citenamefont {Mazilu}, \citenamefont {Milburn},
  \citenamefont {M{\" u}ller}, \citenamefont {Novotny}, \citenamefont
  {Paternostro}, \citenamefont {Peters}, \citenamefont {Pikovski},
  \citenamefont {Pilan~Zanoni}, \citenamefont {Rasel}, \citenamefont {Reynaud},
  \citenamefont {Riedel}, \citenamefont {Rodrigues}, \citenamefont {Rondin},
  \citenamefont {Roura}, \citenamefont {Schleich}, \citenamefont
  {Schmiedmayer}, \citenamefont {Schuldt}, \citenamefont {Schwab},
  \citenamefont {Tajmar}, \citenamefont {Tino}, \citenamefont {Ulbricht},
  \citenamefont {Ursin},\ and\ \citenamefont
  {Vedral}}]{kaltenbaek_macroscopic_2016}%
  \BibitemOpen
  \bibfield  {author} {\bibinfo {author} {\bibfnamefont {Rainer}\ \bibnamefont
  {Kaltenbaek}}, \bibinfo {author} {\bibfnamefont {Markus}\ \bibnamefont
  {Aspelmeyer}}, \bibinfo {author} {\bibfnamefont {Peter~F.}\ \bibnamefont
  {Barker}}, \bibinfo {author} {\bibfnamefont {Angelo}\ \bibnamefont {Bassi}},
  \bibinfo {author} {\bibfnamefont {James}\ \bibnamefont {Bateman}}, \bibinfo
  {author} {\bibfnamefont {Kai}\ \bibnamefont {Bongs}}, \bibinfo {author}
  {\bibfnamefont {Sougato}\ \bibnamefont {Bose}}, \bibinfo {author}
  {\bibfnamefont {Claus}\ \bibnamefont {Braxmaier}}, \bibinfo {author}
  {\bibfnamefont {{\v C}aslav}\ \bibnamefont {Brukner}}, \bibinfo {author}
  {\bibfnamefont {Bruno}\ \bibnamefont {Christophe}}, \bibinfo {author}
  {\bibfnamefont {Michael}\ \bibnamefont {Chwalla}}, \bibinfo {author}
  {\bibfnamefont {Pierre-Fran{\c c}ois}\ \bibnamefont {Cohadon}}, \bibinfo
  {author} {\bibfnamefont {Adrian~Michael}\ \bibnamefont {Cruise}}, \bibinfo
  {author} {\bibfnamefont {Catalina}\ \bibnamefont {Curceanu}}, \bibinfo
  {author} {\bibfnamefont {Kishan}\ \bibnamefont {Dholakia}}, \bibinfo {author}
  {\bibfnamefont {Lajos}\ \bibnamefont {Di{\' o}si}}, \bibinfo {author}
  {\bibfnamefont {Klaus}\ \bibnamefont {D{\" o}ringshoff}}, \bibinfo {author}
  {\bibfnamefont {Wolfgang}\ \bibnamefont {Ertmer}}, \bibinfo {author}
  {\bibfnamefont {Jan}\ \bibnamefont {Gieseler}}, \bibinfo {author}
  {\bibfnamefont {Norman}\ \bibnamefont {G{\" u}rlebeck}}, \bibinfo {author}
  {\bibfnamefont {Gerald}\ \bibnamefont {Hechenblaikner}}, \bibinfo {author}
  {\bibfnamefont {Antoine}\ \bibnamefont {Heidmann}}, \bibinfo {author}
  {\bibfnamefont {Sven}\ \bibnamefont {Herrmann}}, \bibinfo {author}
  {\bibfnamefont {Sabine}\ \bibnamefont {Hossenfelder}}, \bibinfo {author}
  {\bibfnamefont {Ulrich}\ \bibnamefont {Johann}}, \bibinfo {author}
  {\bibfnamefont {Nikolai}\ \bibnamefont {Kiesel}}, \bibinfo {author}
  {\bibfnamefont {Myungshik}\ \bibnamefont {Kim}}, \bibinfo {author}
  {\bibfnamefont {Claus}\ \bibnamefont {L{\" a}mmerzahl}}, \bibinfo {author}
  {\bibfnamefont {Astrid}\ \bibnamefont {Lambrecht}}, \bibinfo {author}
  {\bibfnamefont {Michael}\ \bibnamefont {Mazilu}}, \bibinfo {author}
  {\bibfnamefont {Gerard~J.}\ \bibnamefont {Milburn}}, \bibinfo {author}
  {\bibfnamefont {Holger}\ \bibnamefont {M{\" u}ller}}, \bibinfo {author}
  {\bibfnamefont {Lukas}\ \bibnamefont {Novotny}}, \bibinfo {author}
  {\bibfnamefont {Mauro}\ \bibnamefont {Paternostro}}, \bibinfo {author}
  {\bibfnamefont {Achim}\ \bibnamefont {Peters}}, \bibinfo {author}
  {\bibfnamefont {Igor}\ \bibnamefont {Pikovski}}, \bibinfo {author}
  {\bibfnamefont {Andr{\' e}}\ \bibnamefont {Pilan~Zanoni}}, \bibinfo {author}
  {\bibfnamefont {Ernst~M.}\ \bibnamefont {Rasel}}, \bibinfo {author}
  {\bibfnamefont {Serge}\ \bibnamefont {Reynaud}}, \bibinfo {author}
  {\bibfnamefont {Charles~Jess}\ \bibnamefont {Riedel}}, \bibinfo {author}
  {\bibfnamefont {Manuel}\ \bibnamefont {Rodrigues}}, \bibinfo {author}
  {\bibfnamefont {Lo\"{\i}c}\ \bibnamefont {Rondin}}, \bibinfo {author}
  {\bibfnamefont {Albert}\ \bibnamefont {Roura}}, \bibinfo {author}
  {\bibfnamefont {Wolfgang~P.}\ \bibnamefont {Schleich}}, \bibinfo {author}
  {\bibfnamefont {J{\" o}rg}\ \bibnamefont {Schmiedmayer}}, \bibinfo {author}
  {\bibfnamefont {Thilo}\ \bibnamefont {Schuldt}}, \bibinfo {author}
  {\bibfnamefont {Keith~C.}\ \bibnamefont {Schwab}}, \bibinfo {author}
  {\bibfnamefont {Martin}\ \bibnamefont {Tajmar}}, \bibinfo {author}
  {\bibfnamefont {Guglielmo~M.}\ \bibnamefont {Tino}}, \bibinfo {author}
  {\bibfnamefont {Hendrik}\ \bibnamefont {Ulbricht}}, \bibinfo {author}
  {\bibfnamefont {Rupert}\ \bibnamefont {Ursin}}, \ and\ \bibinfo {author}
  {\bibfnamefont {Vlatko}\ \bibnamefont {Vedral}},\ }\bibfield  {title}
  {\enquote {\bibinfo {title} {Macroscopic {Quantum} {Resonators} ({MAQRO}):
  2015 update},}\ }\href {\doibase 10.1140/epjqt/s40507-016-0043-7} {\bibfield
  {journal} {\bibinfo  {journal} {EPJ Quantum Technology}\ }\textbf {\bibinfo
  {volume} {3}},\ \bibinfo {pages} {5} (\bibinfo {year} {2016})},\ \bibinfo
  {note} {arXiv: 1503.02640}\BibitemShut {NoStop}%
\bibitem [{\citenamefont {Lei}\ \emph {et~al.}(2016)\citenamefont {Lei},
  \citenamefont {Weinstein}, \citenamefont {Suh}, \citenamefont {Wollman},
  \citenamefont {Kronwald}, \citenamefont {Marquardt}, \citenamefont {Clerk},\
  and\ \citenamefont {Schwab}}]{lei_quantum_2016}%
  \BibitemOpen
  \bibfield  {author} {\bibinfo {author} {\bibfnamefont {C.~U.}\ \bibnamefont
  {Lei}}, \bibinfo {author} {\bibfnamefont {A.~J.}\ \bibnamefont {Weinstein}},
  \bibinfo {author} {\bibfnamefont {J.}~\bibnamefont {Suh}}, \bibinfo {author}
  {\bibfnamefont {E.~E.}\ \bibnamefont {Wollman}}, \bibinfo {author}
  {\bibfnamefont {A.}~\bibnamefont {Kronwald}}, \bibinfo {author}
  {\bibfnamefont {F.}~\bibnamefont {Marquardt}}, \bibinfo {author}
  {\bibfnamefont {A.~A.}\ \bibnamefont {Clerk}}, \ and\ \bibinfo {author}
  {\bibfnamefont {K.~C.}\ \bibnamefont {Schwab}},\ }\bibfield  {title}
  {\enquote {\bibinfo {title} {Quantum {Nondemolition} {Measurement} of a
  {Quantum} {Squeezed} {State} {Beyond} the 3 {dB} {Limit}},}\ }\href {\doibase
  10.1103/PhysRevLett.117.100801} {\bibfield  {journal} {\bibinfo  {journal}
  {Physical Review Letters}\ }\textbf {\bibinfo {volume} {117}},\ \bibinfo
  {pages} {100801} (\bibinfo {year} {2016})},\ \bibinfo {note} {arXiv:
  1605.08148}\BibitemShut {NoStop}%
\bibitem [{\citenamefont {Santos}\ \emph {et~al.}(2017)\citenamefont {Santos},
  \citenamefont {Li}, \citenamefont {Ilves}, \citenamefont {Ockeloen-Korppi},\
  and\ \citenamefont {Sillanp{\" a}{\" a}}}]{santos_optomechanical_2017}%
  \BibitemOpen
  \bibfield  {author} {\bibinfo {author} {\bibfnamefont {J.~T.}\ \bibnamefont
  {Santos}}, \bibinfo {author} {\bibfnamefont {J.}~\bibnamefont {Li}}, \bibinfo
  {author} {\bibfnamefont {J.}~\bibnamefont {Ilves}}, \bibinfo {author}
  {\bibfnamefont {C.~F.}\ \bibnamefont {Ockeloen-Korppi}}, \ and\ \bibinfo
  {author} {\bibfnamefont {M.}~\bibnamefont {Sillanp{\" a}{\" a}}},\ }\bibfield
   {title} {\enquote {\bibinfo {title} {Optomechanical measurement of a
  millimeter-sized mechanical oscillator approaching the quantum ground
  state},}\ }\href {\doibase 10.1088/1367-2630/aa83a5} {\bibfield  {journal}
  {\bibinfo  {journal} {New Journal of Physics}\ }\textbf {\bibinfo {volume}
  {19}},\ \bibinfo {pages} {103014} (\bibinfo {year} {2017})}\BibitemShut
  {NoStop}%
\bibitem [{\citenamefont {Romero-Isart}(2011)}]{romero-isart_quantum_2011}%
  \BibitemOpen
  \bibfield  {author} {\bibinfo {author} {\bibfnamefont {Oriol}\ \bibnamefont
  {Romero-Isart}},\ }\bibfield  {title} {\enquote {\bibinfo {title} {Quantum
  superposition of massive objects and collapse models},}\ }\href {\doibase
  10.1103/PhysRevA.84.052121} {\bibfield  {journal} {\bibinfo  {journal}
  {Physical Review A}\ }\textbf {\bibinfo {volume} {84}},\ \bibinfo {pages}
  {052121} (\bibinfo {year} {2011})}\BibitemShut {NoStop}%
\bibitem [{\citenamefont {Bera}\ \emph {et~al.}(2015)\citenamefont {Bera},
  \citenamefont {Motwani}, \citenamefont {Singh},\ and\ \citenamefont
  {Ulbricht}}]{bera_proposal_2015}%
  \BibitemOpen
  \bibfield  {author} {\bibinfo {author} {\bibfnamefont {Sayantani}\
  \bibnamefont {Bera}}, \bibinfo {author} {\bibfnamefont {Bhawna}\ \bibnamefont
  {Motwani}}, \bibinfo {author} {\bibfnamefont {Tejinder~P.}\ \bibnamefont
  {Singh}}, \ and\ \bibinfo {author} {\bibfnamefont {Hendrik}\ \bibnamefont
  {Ulbricht}},\ }\bibfield  {title} {\enquote {\bibinfo {title} {A proposal for
  the experimental detection of {CSL} induced random walk},}\ }\href {\doibase
  10.1038/srep07664} {\bibfield  {journal} {\bibinfo  {journal} {Scientific
  Reports}\ }\textbf {\bibinfo {volume} {5}},\ \bibinfo {pages} {7664}
  (\bibinfo {year} {2015})}\BibitemShut {NoStop}%
\bibitem [{\citenamefont {Vinante}\ \emph {et~al.}(2017)\citenamefont
  {Vinante}, \citenamefont {Mezzena}, \citenamefont {Falferi}, \citenamefont
  {Carlesso},\ and\ \citenamefont {Bassi}}]{vinante_improved_2017}%
  \BibitemOpen
  \bibfield  {author} {\bibinfo {author} {\bibfnamefont {A.}~\bibnamefont
  {Vinante}}, \bibinfo {author} {\bibfnamefont {R.}~\bibnamefont {Mezzena}},
  \bibinfo {author} {\bibfnamefont {P.}~\bibnamefont {Falferi}}, \bibinfo
  {author} {\bibfnamefont {M.}~\bibnamefont {Carlesso}}, \ and\ \bibinfo
  {author} {\bibfnamefont {A.}~\bibnamefont {Bassi}},\ }\bibfield  {title}
  {\enquote {\bibinfo {title} {Improved {Noninterferometric} {Test} of
  {Collapse} {Models} {Using} {Ultracold} {Cantilevers}},}\ }\href {\doibase
  10.1103/PhysRevLett.119.110401} {\bibfield  {journal} {\bibinfo  {journal}
  {Physical Review Letters}\ }\textbf {\bibinfo {volume} {119}},\ \bibinfo
  {pages} {110401} (\bibinfo {year} {2017})}\BibitemShut {NoStop}%
\bibitem [{\citenamefont {Chang}\ \emph {et~al.}(2010)\citenamefont {Chang},
  \citenamefont {Regal}, \citenamefont {Papp}, \citenamefont {Wilson},
  \citenamefont {Ye}, \citenamefont {Painter}, \citenamefont {Kimble},\ and\
  \citenamefont {Zoller}}]{chang_cavity_2010}%
  \BibitemOpen
  \bibfield  {author} {\bibinfo {author} {\bibfnamefont {D.~E.}\ \bibnamefont
  {Chang}}, \bibinfo {author} {\bibfnamefont {C.~A.}\ \bibnamefont {Regal}},
  \bibinfo {author} {\bibfnamefont {S.~B.}\ \bibnamefont {Papp}}, \bibinfo
  {author} {\bibfnamefont {D.~J.}\ \bibnamefont {Wilson}}, \bibinfo {author}
  {\bibfnamefont {J.}~\bibnamefont {Ye}}, \bibinfo {author} {\bibfnamefont
  {O.}~\bibnamefont {Painter}}, \bibinfo {author} {\bibfnamefont {H.~J.}\
  \bibnamefont {Kimble}}, \ and\ \bibinfo {author} {\bibfnamefont
  {P.}~\bibnamefont {Zoller}},\ }\bibfield  {title} {\enquote {\bibinfo {title}
  {Cavity opto-mechanics using an optically levitated nanosphere},}\ }\href
  {\doibase 10.1073/pnas.0912969107} {\bibfield  {journal} {\bibinfo  {journal}
  {Proceedings of the National Academy of Sciences}\ }\textbf {\bibinfo
  {volume} {107}},\ \bibinfo {pages} {1005--1010} (\bibinfo {year} {2010})},\
  \bibinfo {note} {arXiv: 0909.1548}\BibitemShut {NoStop}%
\bibitem [{\citenamefont {Romero-Isart}\ \emph {et~al.}(2011)\citenamefont
  {Romero-Isart}, \citenamefont {Pflanzer}, \citenamefont {Juan}, \citenamefont
  {Quidant}, \citenamefont {Kiesel}, \citenamefont {Aspelmeyer},\ and\
  \citenamefont {Cirac}}]{romero-isart_optically_2011}%
  \BibitemOpen
  \bibfield  {author} {\bibinfo {author} {\bibfnamefont {O.}~\bibnamefont
  {Romero-Isart}}, \bibinfo {author} {\bibfnamefont {A.~C.}\ \bibnamefont
  {Pflanzer}}, \bibinfo {author} {\bibfnamefont {M.~L.}\ \bibnamefont {Juan}},
  \bibinfo {author} {\bibfnamefont {R.}~\bibnamefont {Quidant}}, \bibinfo
  {author} {\bibfnamefont {N.}~\bibnamefont {Kiesel}}, \bibinfo {author}
  {\bibfnamefont {M.}~\bibnamefont {Aspelmeyer}}, \ and\ \bibinfo {author}
  {\bibfnamefont {J.~I.}\ \bibnamefont {Cirac}},\ }\bibfield  {title} {\enquote
  {\bibinfo {title} {Optically levitating dielectrics in the quantum regime:
  {Theory} and protocols},}\ }\href {\doibase 10.1103/PhysRevA.83.013803}
  {\bibfield  {journal} {\bibinfo  {journal} {Physical Review A}\ }\textbf
  {\bibinfo {volume} {83}},\ \bibinfo {pages} {013803} (\bibinfo {year}
  {2011})},\ \bibinfo {note} {arXiv: 1010.3109}\BibitemShut {NoStop}%
\bibitem [{\citenamefont {Barker}(2010)}]{barker_doppler_2010}%
  \BibitemOpen
  \bibfield  {author} {\bibinfo {author} {\bibfnamefont {P.~F.}\ \bibnamefont
  {Barker}},\ }\bibfield  {title} {\enquote {\bibinfo {title} {Doppler
  {Cooling} a {Microsphere}},}\ }\href {\doibase
  10.1103/PhysRevLett.105.073002} {\bibfield  {journal} {\bibinfo  {journal}
  {Physical Review Letters}\ }\textbf {\bibinfo {volume} {105}},\ \bibinfo
  {pages} {073002} (\bibinfo {year} {2010})}\BibitemShut {NoStop}%
\bibitem [{\citenamefont {Kiesel}\ \emph {et~al.}(2013)\citenamefont {Kiesel},
  \citenamefont {Blaser}, \citenamefont {Deli{\' c}}, \citenamefont {Grass},
  \citenamefont {Kaltenbaek},\ and\ \citenamefont
  {Aspelmeyer}}]{kiesel_cavity_2013}%
  \BibitemOpen
  \bibfield  {author} {\bibinfo {author} {\bibfnamefont {Nikolai}\ \bibnamefont
  {Kiesel}}, \bibinfo {author} {\bibfnamefont {Florian}\ \bibnamefont
  {Blaser}}, \bibinfo {author} {\bibfnamefont {Uro{\v s}}\ \bibnamefont
  {Deli{\' c}}}, \bibinfo {author} {\bibfnamefont {David}\ \bibnamefont
  {Grass}}, \bibinfo {author} {\bibfnamefont {Rainer}\ \bibnamefont
  {Kaltenbaek}}, \ and\ \bibinfo {author} {\bibfnamefont {Markus}\ \bibnamefont
  {Aspelmeyer}},\ }\bibfield  {title} {\enquote {\bibinfo {title} {Cavity
  cooling of an optically levitated submicron particle},}\ }\href {\doibase
  10.1073/pnas.1309167110} {\bibfield  {journal} {\bibinfo  {journal}
  {Proceedings of the National Academy of Sciences}\ }\textbf {\bibinfo
  {volume} {110}},\ \bibinfo {pages} {14180--14185} (\bibinfo {year} {2013})},\
  \bibinfo {note} {arXiv: 1304.6679}\BibitemShut {NoStop}%
\bibitem [{\citenamefont {Yin}\ \emph {et~al.}(2013)\citenamefont {Yin},
  \citenamefont {Geraci},\ and\ \citenamefont {Li}}]{yin_optomechanics_2013}%
  \BibitemOpen
  \bibfield  {author} {\bibinfo {author} {\bibfnamefont {Zhang-Qi}\
  \bibnamefont {Yin}}, \bibinfo {author} {\bibfnamefont {Andrew~A.}\
  \bibnamefont {Geraci}}, \ and\ \bibinfo {author} {\bibfnamefont {Tongcang}\
  \bibnamefont {Li}},\ }\bibfield  {title} {\enquote {\bibinfo {title}
  {Optomechanics of levitated dielectric particles},}\ }\href {\doibase
  10.1142/S0217979213300181} {\bibfield  {journal} {\bibinfo  {journal}
  {International Journal of Modern Physics B}\ }\textbf {\bibinfo {volume}
  {27}},\ \bibinfo {pages} {1330018} (\bibinfo {year} {2013})}\BibitemShut
  {NoStop}%
\bibitem [{\citenamefont {Dholakia}\ and\ \citenamefont {{\v C}i{\v z}m{\'
  a}r}(2011)}]{dholakia_shaping_2011}%
  \BibitemOpen
  \bibfield  {author} {\bibinfo {author} {\bibfnamefont {K.}~\bibnamefont
  {Dholakia}}\ and\ \bibinfo {author} {\bibfnamefont {T.}~\bibnamefont {{\v
  C}i{\v z}m{\' a}r}},\ }\bibfield  {title} {\enquote {\bibinfo {title}
  {Shaping the future of manipulation},}\ }\href {\doibase
  10.1038/nphoton.2011.80} {\bibfield  {journal} {\bibinfo  {journal} {Nature
  Photonics}\ }\textbf {\bibinfo {volume} {5}},\ \bibinfo {pages} {335--342}
  (\bibinfo {year} {2011})}\BibitemShut {NoStop}%
\bibitem [{\citenamefont {Gieseler}\ \emph {et~al.}(2013)\citenamefont
  {Gieseler}, \citenamefont {Novotny},\ and\ \citenamefont
  {Quidant}}]{gieseler_thermal_2013}%
  \BibitemOpen
  \bibfield  {author} {\bibinfo {author} {\bibfnamefont {Jan}\ \bibnamefont
  {Gieseler}}, \bibinfo {author} {\bibfnamefont {Lukas}\ \bibnamefont
  {Novotny}}, \ and\ \bibinfo {author} {\bibfnamefont {Romain}\ \bibnamefont
  {Quidant}},\ }\bibfield  {title} {\enquote {\bibinfo {title} {Thermal
  nonlinearities in a nanomechanical oscillator},}\ }\href {\doibase
  10.1038/nphys2798} {\bibfield  {journal} {\bibinfo  {journal} {Nature
  Physics}\ }\textbf {\bibinfo {volume} {9}},\ \bibinfo {pages} {806--810}
  (\bibinfo {year} {2013})}\BibitemShut {NoStop}%
\bibitem [{\citenamefont {Fonseca}\ \emph {et~al.}(2016)\citenamefont
  {Fonseca}, \citenamefont {Aranas}, \citenamefont {Millen}, \citenamefont
  {Monteiro},\ and\ \citenamefont {Barker}}]{fonseca_nonlinear_2016}%
  \BibitemOpen
  \bibfield  {author} {\bibinfo {author} {\bibfnamefont {P.~Z.~G.}\
  \bibnamefont {Fonseca}}, \bibinfo {author} {\bibfnamefont {E.~B.}\
  \bibnamefont {Aranas}}, \bibinfo {author} {\bibfnamefont {J.}~\bibnamefont
  {Millen}}, \bibinfo {author} {\bibfnamefont {T.~S.}\ \bibnamefont
  {Monteiro}}, \ and\ \bibinfo {author} {\bibfnamefont {P.~F.}\ \bibnamefont
  {Barker}},\ }\bibfield  {title} {\enquote {\bibinfo {title} {Nonlinear
  {Dynamics} and {Strong} {Cavity} {Cooling} of {Levitated} {Nanoparticles}},}\
  }\href {\doibase 10.1103/PhysRevLett.117.173602} {\bibfield  {journal}
  {\bibinfo  {journal} {Physical Review Letters}\ }\textbf {\bibinfo {volume}
  {117}},\ \bibinfo {pages} {173602} (\bibinfo {year} {2016})}\BibitemShut
  {NoStop}%
\bibitem [{\citenamefont {Ricci}\ \emph {et~al.}(2017)\citenamefont {Ricci},
  \citenamefont {Rica}, \citenamefont {Spasenovi{\' c}}, \citenamefont
  {Gieseler}, \citenamefont {Rondin}, \citenamefont {Novotny},\ and\
  \citenamefont {Quidant}}]{ricci_optically_2017}%
  \BibitemOpen
  \bibfield  {author} {\bibinfo {author} {\bibfnamefont {F.}~\bibnamefont
  {Ricci}}, \bibinfo {author} {\bibfnamefont {R.~A.}\ \bibnamefont {Rica}},
  \bibinfo {author} {\bibfnamefont {M.}~\bibnamefont {Spasenovi{\' c}}},
  \bibinfo {author} {\bibfnamefont {J.}~\bibnamefont {Gieseler}}, \bibinfo
  {author} {\bibfnamefont {L.}~\bibnamefont {Rondin}}, \bibinfo {author}
  {\bibfnamefont {L.}~\bibnamefont {Novotny}}, \ and\ \bibinfo {author}
  {\bibfnamefont {R.}~\bibnamefont {Quidant}},\ }\bibfield  {title} {\enquote
  {\bibinfo {title} {Optically levitated nanoparticle as a model system for
  stochastic bistable dynamics},}\ }\href {\doibase 10.1038/ncomms15141}
  {\bibfield  {journal} {\bibinfo  {journal} {Nature Communications}\ }\textbf
  {\bibinfo {volume} {8}},\ \bibinfo {pages} {15141} (\bibinfo {year}
  {2017})},\ \bibinfo {note} {arXiv: 1705.04061}\BibitemShut {NoStop}%
\bibitem [{\citenamefont {{\v S}iler}\ \emph {et~al.}(2017)\citenamefont {{\v
  S}iler}, \citenamefont {J{\' a}kl}, \citenamefont {Brzobohat{\' y}},
  \citenamefont {Ryabov}, \citenamefont {Filip},\ and\ \citenamefont {Zem{\'
  a}nek}}]{siler_thermally_2017}%
  \BibitemOpen
  \bibfield  {author} {\bibinfo {author} {\bibfnamefont {Martin}\ \bibnamefont
  {{\v S}iler}}, \bibinfo {author} {\bibfnamefont {Petr}\ \bibnamefont {J{\'
  a}kl}}, \bibinfo {author} {\bibfnamefont {Oto}\ \bibnamefont {Brzobohat{\'
  y}}}, \bibinfo {author} {\bibfnamefont {Artem}\ \bibnamefont {Ryabov}},
  \bibinfo {author} {\bibfnamefont {Radim}\ \bibnamefont {Filip}}, \ and\
  \bibinfo {author} {\bibfnamefont {Pavel}\ \bibnamefont {Zem{\' a}nek}},\
  }\bibfield  {title} {\enquote {\bibinfo {title} {Thermally induced
  micro-motion by inflection in optical potential},}\ }\href {\doibase
  10.1038/s41598-017-01848-4} {\bibfield  {journal} {\bibinfo  {journal}
  {Scientific Reports}\ }\textbf {\bibinfo {volume} {7}},\ \bibinfo {pages}
  {1697} (\bibinfo {year} {2017})}\BibitemShut {NoStop}%
\bibitem [{\citenamefont {{\v S}iler}\ \emph {et~al.}(2018)\citenamefont {{\v
  S}iler}, \citenamefont {Ornigotti}, \citenamefont {Brzobohat{\' y}},
  \citenamefont {J{\' a}kl}, \citenamefont {Ryabov}, \citenamefont {Holubec},
  \citenamefont {Zem{\' a}nek},\ and\ \citenamefont
  {Filip}}]{siler_diffusing_2018}%
  \BibitemOpen
  \bibfield  {author} {\bibinfo {author} {\bibfnamefont {Martin}\ \bibnamefont
  {{\v S}iler}}, \bibinfo {author} {\bibfnamefont {Luca}\ \bibnamefont
  {Ornigotti}}, \bibinfo {author} {\bibfnamefont {Oto}\ \bibnamefont
  {Brzobohat{\' y}}}, \bibinfo {author} {\bibfnamefont {Petr}\ \bibnamefont
  {J{\' a}kl}}, \bibinfo {author} {\bibfnamefont {Artem}\ \bibnamefont
  {Ryabov}}, \bibinfo {author} {\bibfnamefont {Viktor}\ \bibnamefont
  {Holubec}}, \bibinfo {author} {\bibfnamefont {Pavel}\ \bibnamefont {Zem{\'
  a}nek}}, \ and\ \bibinfo {author} {\bibfnamefont {Radim}\ \bibnamefont
  {Filip}},\ }\bibfield  {title} {\enquote {\bibinfo {title} {Diffusing {Up}
  the {Hill}: {Dynamics} and {Equipartition} in {Highly} {Unstable}
  {Systems}},}\ }\href@noop {} {\bibfield  {journal} {\bibinfo  {journal}
  {arXiv:1803.07833 [cond-mat]}\ } (\bibinfo {year} {2018})},\ \bibinfo {note}
  {arXiv: 1803.07833}\BibitemShut {NoStop}%
\bibitem [{\citenamefont {Konopik}\ \emph {et~al.}(2018)\citenamefont
  {Konopik}, \citenamefont {Friedenberger}, \citenamefont {Kiesel},\ and\
  \citenamefont {Lutz}}]{konopik_nonequilibrium_2018}%
  \BibitemOpen
  \bibfield  {author} {\bibinfo {author} {\bibfnamefont {Michael}\ \bibnamefont
  {Konopik}}, \bibinfo {author} {\bibfnamefont {Alexander}\ \bibnamefont
  {Friedenberger}}, \bibinfo {author} {\bibfnamefont {Nikolai}\ \bibnamefont
  {Kiesel}}, \ and\ \bibinfo {author} {\bibfnamefont {Eric}\ \bibnamefont
  {Lutz}},\ }\bibfield  {title} {\enquote {\bibinfo {title} {Nonequilibrium
  information erasure below {kTln}2},}\ }\href@noop {} {\bibfield  {journal}
  {\bibinfo  {journal} {arXiv:1806.01034 [cond-mat]}\ } (\bibinfo {year}
  {2018})},\ \bibinfo {note} {arXiv: 1806.01034}\BibitemShut {NoStop}%
\bibitem [{\citenamefont {Millen}\ \emph {et~al.}(2015)\citenamefont {Millen},
  \citenamefont {Fonseca}, \citenamefont {Mavrogordatos}, \citenamefont
  {Monteiro},\ and\ \citenamefont {Barker}}]{millen_cavity_2015}%
  \BibitemOpen
  \bibfield  {author} {\bibinfo {author} {\bibfnamefont {J.}~\bibnamefont
  {Millen}}, \bibinfo {author} {\bibfnamefont {P.~Z.~G.}\ \bibnamefont
  {Fonseca}}, \bibinfo {author} {\bibfnamefont {T.}~\bibnamefont
  {Mavrogordatos}}, \bibinfo {author} {\bibfnamefont {T.~S.}\ \bibnamefont
  {Monteiro}}, \ and\ \bibinfo {author} {\bibfnamefont {P.~F.}\ \bibnamefont
  {Barker}},\ }\bibfield  {title} {\enquote {\bibinfo {title} {Cavity {Cooling}
  a {Single} {Charged} {Levitated} {Nanosphere}},}\ }\href {\doibase
  10.1103/PhysRevLett.114.123602} {\bibfield  {journal} {\bibinfo  {journal}
  {Physical Review Letters}\ }\textbf {\bibinfo {volume} {114}},\ \bibinfo
  {pages} {123602} (\bibinfo {year} {2015})}\BibitemShut {NoStop}%
\bibitem [{\citenamefont {Gieseler}\ \emph {et~al.}(2012)\citenamefont
  {Gieseler}, \citenamefont {Deutsch}, \citenamefont {Quidant},\ and\
  \citenamefont {Novotny}}]{gieseler_subkelvin_2012}%
  \BibitemOpen
  \bibfield  {author} {\bibinfo {author} {\bibfnamefont {Jan}\ \bibnamefont
  {Gieseler}}, \bibinfo {author} {\bibfnamefont {Bradley}\ \bibnamefont
  {Deutsch}}, \bibinfo {author} {\bibfnamefont {Romain}\ \bibnamefont
  {Quidant}}, \ and\ \bibinfo {author} {\bibfnamefont {Lukas}\ \bibnamefont
  {Novotny}},\ }\bibfield  {title} {\enquote {\bibinfo {title} {Subkelvin
  {Parametric} {Feedback} {Cooling} of a {Laser}-{Trapped} {Nanoparticle}},}\
  }\href {\doibase 10.1103/PhysRevLett.109.103603} {\bibfield  {journal}
  {\bibinfo  {journal} {Physical Review Letters}\ }\textbf {\bibinfo {volume}
  {109}},\ \bibinfo {pages} {103603} (\bibinfo {year} {2012})}\BibitemShut
  {NoStop}%
\bibitem [{\citenamefont {Asenbaum}\ \emph {et~al.}(2013)\citenamefont
  {Asenbaum}, \citenamefont {Kuhn}, \citenamefont {Nimmrichter}, \citenamefont
  {Sezer},\ and\ \citenamefont {Arndt}}]{asenbaum_cavity_2013}%
  \BibitemOpen
  \bibfield  {author} {\bibinfo {author} {\bibfnamefont {Peter}\ \bibnamefont
  {Asenbaum}}, \bibinfo {author} {\bibfnamefont {Stefan}\ \bibnamefont {Kuhn}},
  \bibinfo {author} {\bibfnamefont {Stefan}\ \bibnamefont {Nimmrichter}},
  \bibinfo {author} {\bibfnamefont {Ugur}\ \bibnamefont {Sezer}}, \ and\
  \bibinfo {author} {\bibfnamefont {Markus}\ \bibnamefont {Arndt}},\ }\bibfield
   {title} {\enquote {\bibinfo {title} {Cavity cooling of free silicon
  nanoparticles in high vacuum},}\ }\href {\doibase 10.1038/ncomms3743}
  {\bibfield  {journal} {\bibinfo  {journal} {Nature Communications}\ }\textbf
  {\bibinfo {volume} {4}},\ \bibinfo {pages} {2743} (\bibinfo {year} {2013})},\
  \bibinfo {note} {arXiv: 1306.4617}\BibitemShut {NoStop}%
\bibitem [{\citenamefont {Jain}\ \emph {et~al.}(2016)\citenamefont {Jain},
  \citenamefont {Gieseler}, \citenamefont {Moritz}, \citenamefont {Dellago},
  \citenamefont {Quidant},\ and\ \citenamefont {Novotny}}]{jain_direct_2016}%
  \BibitemOpen
  \bibfield  {author} {\bibinfo {author} {\bibfnamefont {Vijay}\ \bibnamefont
  {Jain}}, \bibinfo {author} {\bibfnamefont {Jan}\ \bibnamefont {Gieseler}},
  \bibinfo {author} {\bibfnamefont {Clemens}\ \bibnamefont {Moritz}}, \bibinfo
  {author} {\bibfnamefont {Christoph}\ \bibnamefont {Dellago}}, \bibinfo
  {author} {\bibfnamefont {Romain}\ \bibnamefont {Quidant}}, \ and\ \bibinfo
  {author} {\bibfnamefont {Lukas}\ \bibnamefont {Novotny}},\ }\bibfield
  {title} {\enquote {\bibinfo {title} {Direct {Measurement} of {Photon}
  {Recoil} from a {Levitated} {Nanoparticle}},}\ }\href {\doibase
  10.1103/PhysRevLett.116.243601} {\bibfield  {journal} {\bibinfo  {journal}
  {Physical Review Letters}\ }\textbf {\bibinfo {volume} {116}},\ \bibinfo
  {pages} {243601} (\bibinfo {year} {2016})},\ \bibinfo {note} {arXiv:
  1603.03420}\BibitemShut {NoStop}%
\bibitem [{\citenamefont {Rashid}\ \emph {et~al.}(2016)\citenamefont {Rashid},
  \citenamefont {Tufarelli}, \citenamefont {Bateman}, \citenamefont {Vovrosh},
  \citenamefont {Hempston}, \citenamefont {Kim},\ and\ \citenamefont
  {Ulbricht}}]{rashid_experimental_2016}%
  \BibitemOpen
  \bibfield  {author} {\bibinfo {author} {\bibfnamefont {Muddassar}\
  \bibnamefont {Rashid}}, \bibinfo {author} {\bibfnamefont {Tommaso}\
  \bibnamefont {Tufarelli}}, \bibinfo {author} {\bibfnamefont {James}\
  \bibnamefont {Bateman}}, \bibinfo {author} {\bibfnamefont {Jamie}\
  \bibnamefont {Vovrosh}}, \bibinfo {author} {\bibfnamefont {David}\
  \bibnamefont {Hempston}}, \bibinfo {author} {\bibfnamefont {M.~S.}\
  \bibnamefont {Kim}}, \ and\ \bibinfo {author} {\bibfnamefont {Hendrik}\
  \bibnamefont {Ulbricht}},\ }\bibfield  {title} {\enquote {\bibinfo {title}
  {Experimental {Realization} of a {Thermal} {Squeezed} {State} of {Levitated}
  {Optomechanics}},}\ }\href {\doibase 10.1103/PhysRevLett.117.273601}
  {\bibfield  {journal} {\bibinfo  {journal} {Physical Review Letters}\
  }\textbf {\bibinfo {volume} {117}},\ \bibinfo {pages} {273601} (\bibinfo
  {year} {2016})},\ \bibinfo {note} {arXiv: 1607.05509}\BibitemShut {NoStop}%
\bibitem [{\citenamefont {Bowen}\ and\ \citenamefont
  {Milburn}(2015)}]{bowen_quantum_2015}%
  \BibitemOpen
  \bibfield  {author} {\bibinfo {author} {\bibfnamefont {Warwick~P.}\
  \bibnamefont {Bowen}}\ and\ \bibinfo {author} {\bibfnamefont {Gerard~J.}\
  \bibnamefont {Milburn}},\ }\href@noop {} {\emph {\bibinfo {title} {Quantum
  {Optomechanics}}}}\ (\bibinfo  {publisher} {CRC Press},\ \bibinfo {year}
  {2015})\ \bibinfo {note} {google-Books-ID: YZDwCgAAQBAJ}\BibitemShut
  {NoStop}%
\bibitem [{\citenamefont {Law}(1995)}]{law_interaction_1995}%
  \BibitemOpen
  \bibfield  {author} {\bibinfo {author} {\bibfnamefont {C.~K.}\ \bibnamefont
  {Law}},\ }\bibfield  {title} {\enquote {\bibinfo {title} {Interaction between
  a moving mirror and radiation pressure: A {Hamiltonian} formulation},}\
  }\href {\doibase 10.1103/PhysRevA.51.2537} {\bibfield  {journal} {\bibinfo
  {journal} {Physical Review A}\ }\textbf {\bibinfo {volume} {51}},\ \bibinfo
  {pages} {2537--2541} (\bibinfo {year} {1995})}\BibitemShut {NoStop}%
\bibitem [{\citenamefont {Monteiro}\ \emph {et~al.}(2013)\citenamefont
  {Monteiro}, \citenamefont {Millen}, \citenamefont {Pender}, \citenamefont
  {Marquardt}, \citenamefont {Chang},\ and\ \citenamefont
  {Barker}}]{monteiro_dynamics_2013}%
  \BibitemOpen
  \bibfield  {author} {\bibinfo {author} {\bibfnamefont {T.~S.}\ \bibnamefont
  {Monteiro}}, \bibinfo {author} {\bibfnamefont {J.}~\bibnamefont {Millen}},
  \bibinfo {author} {\bibfnamefont {G.~A.~T.}\ \bibnamefont {Pender}}, \bibinfo
  {author} {\bibfnamefont {Florian}\ \bibnamefont {Marquardt}}, \bibinfo
  {author} {\bibfnamefont {D.}~\bibnamefont {Chang}}, \ and\ \bibinfo {author}
  {\bibfnamefont {P.~F.}\ \bibnamefont {Barker}},\ }\bibfield  {title}
  {\enquote {\bibinfo {title} {Dynamics of levitated nanospheres: towards the
  strong coupling regime},}\ }\href {\doibase 10.1088/1367-2630/15/1/015001}
  {\bibfield  {journal} {\bibinfo  {journal} {New Journal of Physics}\ }\textbf
  {\bibinfo {volume} {15}},\ \bibinfo {pages} {015001} (\bibinfo {year}
  {2013})}\BibitemShut {NoStop}%
\bibitem [{\citenamefont {Filip}\ and\ \citenamefont {Kup{\v
  c}\'{\i}k}(2013)}]{filip_robust_2013}%
  \BibitemOpen
  \bibfield  {author} {\bibinfo {author} {\bibfnamefont {Radim}\ \bibnamefont
  {Filip}}\ and\ \bibinfo {author} {\bibfnamefont {Vojt{\v e}ch}\ \bibnamefont
  {Kup{\v c}\'{\i}k}},\ }\bibfield  {title} {\enquote {\bibinfo {title} {Robust
  {Gaussian} entanglement with a macroscopic oscillator at thermal
  equilibrium},}\ }\href {\doibase 10.1103/PhysRevA.87.062323} {\bibfield
  {journal} {\bibinfo  {journal} {Physical Review A}\ }\textbf {\bibinfo
  {volume} {87}},\ \bibinfo {pages} {062323} (\bibinfo {year}
  {2013})}\BibitemShut {NoStop}%
\bibitem [{\citenamefont {Rakhubovsky}\ and\ \citenamefont
  {Filip}(2015)}]{rakhubovsky_robust_2015}%
  \BibitemOpen
  \bibfield  {author} {\bibinfo {author} {\bibfnamefont {Andrey~A.}\
  \bibnamefont {Rakhubovsky}}\ and\ \bibinfo {author} {\bibfnamefont {Radim}\
  \bibnamefont {Filip}},\ }\bibfield  {title} {\enquote {\bibinfo {title}
  {Robust entanglement with a thermal mechanical oscillator},}\ }\href
  {\doibase 10.1103/PhysRevA.91.062317} {\bibfield  {journal} {\bibinfo
  {journal} {Physical Review A}\ }\textbf {\bibinfo {volume} {91}},\ \bibinfo
  {pages} {062317} (\bibinfo {year} {2015})}\BibitemShut {NoStop}%
\bibitem [{\citenamefont {Vanner}\ \emph {et~al.}(2013)\citenamefont {Vanner},
  \citenamefont {Hofer}, \citenamefont {Cole},\ and\ \citenamefont
  {Aspelmeyer}}]{vanner_cooling-by-measurement_2013}%
  \BibitemOpen
  \bibfield  {author} {\bibinfo {author} {\bibfnamefont {M.~R.}\ \bibnamefont
  {Vanner}}, \bibinfo {author} {\bibfnamefont {J.}~\bibnamefont {Hofer}},
  \bibinfo {author} {\bibfnamefont {G.~D.}\ \bibnamefont {Cole}}, \ and\
  \bibinfo {author} {\bibfnamefont {M.}~\bibnamefont {Aspelmeyer}},\ }\bibfield
   {title} {\enquote {\bibinfo {title} {Cooling-by-measurement and mechanical
  state tomography via pulsed optomechanics},}\ }\href {\doibase
  10.1038/ncomms3295} {\bibfield  {journal} {\bibinfo  {journal} {Nature
  Communications}\ }\textbf {\bibinfo {volume} {4}},\ \bibinfo {pages} {2295}
  (\bibinfo {year} {2013})},\ \bibinfo {note} {arXiv: 1211.7036}\BibitemShut
  {NoStop}%
\bibitem [{\citenamefont {Genes}\ \emph {et~al.}(2009)\citenamefont {Genes},
  \citenamefont {Mari}, \citenamefont {Vitali},\ and\ \citenamefont
  {Tombesi}}]{genes_quantum_2009}%
  \BibitemOpen
  \bibfield  {author} {\bibinfo {author} {\bibfnamefont {C.}~\bibnamefont
  {Genes}}, \bibinfo {author} {\bibfnamefont {A.}~\bibnamefont {Mari}},
  \bibinfo {author} {\bibfnamefont {D.}~\bibnamefont {Vitali}}, \ and\ \bibinfo
  {author} {\bibfnamefont {P.}~\bibnamefont {Tombesi}},\ }\bibfield  {title}
  {\enquote {\bibinfo {title} {Quantum {Effects} in {Optomechanical}
  {Systems}},}\ }in\ \href@noop {} {\emph {\bibinfo {booktitle} {Advances {In}
  {Atomic}, {Molecular}, and {Optical} {Physics}}}},\ Vol.~\bibinfo {volume}
  {57},\ \bibinfo {editor} {edited by\ \bibinfo {editor} {\bibnamefont {{Ennio
  Arimondo}}}, \bibinfo {editor} {\bibnamefont {{Paul R. Berman}}}, \ and\
  \bibinfo {editor} {\bibnamefont {{C. C. Lin}}}}\ (\bibinfo  {publisher}
  {Academic Press},\ \bibinfo {year} {2009})\ pp.\ \bibinfo {pages} {33--86},\
  \bibinfo {note} {http://arxiv.org/abs/0901.2726}\BibitemShut {NoStop}%
\bibitem [{\citenamefont {Giovannetti}\ and\ \citenamefont
  {Vitali}(2001)}]{giovannetti_phase-noise_2001}%
  \BibitemOpen
  \bibfield  {author} {\bibinfo {author} {\bibfnamefont {Vittorio}\
  \bibnamefont {Giovannetti}}\ and\ \bibinfo {author} {\bibfnamefont {David}\
  \bibnamefont {Vitali}},\ }\bibfield  {title} {\enquote {\bibinfo {title}
  {Phase-noise measurement in a cavity with a movable mirror undergoing quantum
  {Brownian} motion},}\ }\href {\doibase 10.1103/PhysRevA.63.023812} {\bibfield
   {journal} {\bibinfo  {journal} {Physical Review A}\ }\textbf {\bibinfo
  {volume} {63}},\ \bibinfo {pages} {023812} (\bibinfo {year}
  {2001})}\BibitemShut {NoStop}%
\bibitem [{\citenamefont {Weedbrook}\ \emph {et~al.}(2012)\citenamefont
  {Weedbrook}, \citenamefont {Pirandola}, \citenamefont {Garc\'{\i}a-Patr{\'
  o}n}, \citenamefont {Cerf}, \citenamefont {Ralph}, \citenamefont {Shapiro},\
  and\ \citenamefont {Lloyd}}]{weedbrook_gaussian_2012}%
  \BibitemOpen
  \bibfield  {author} {\bibinfo {author} {\bibfnamefont {Christian}\
  \bibnamefont {Weedbrook}}, \bibinfo {author} {\bibfnamefont {Stefano}\
  \bibnamefont {Pirandola}}, \bibinfo {author} {\bibfnamefont {Ra{\' u}l}\
  \bibnamefont {Garc\'{\i}a-Patr{\' o}n}}, \bibinfo {author} {\bibfnamefont
  {Nicolas~J.}\ \bibnamefont {Cerf}}, \bibinfo {author} {\bibfnamefont
  {Timothy~C.}\ \bibnamefont {Ralph}}, \bibinfo {author} {\bibfnamefont
  {Jeffrey~H.}\ \bibnamefont {Shapiro}}, \ and\ \bibinfo {author}
  {\bibfnamefont {Seth}\ \bibnamefont {Lloyd}},\ }\bibfield  {title} {\enquote
  {\bibinfo {title} {Gaussian quantum information},}\ }\href {\doibase
  10.1103/RevModPhys.84.621} {\bibfield  {journal} {\bibinfo  {journal}
  {Reviews of Modern Physics}\ }\textbf {\bibinfo {volume} {84}},\ \bibinfo
  {pages} {621--669} (\bibinfo {year} {2012})},\ \bibinfo {note} {arXiv:
  1110.3234}\BibitemShut {NoStop}%
\bibitem [{\citenamefont {Deli{\' c}}\ \emph {et~al.}(2018)\citenamefont
  {Deli{\' c}}, \citenamefont {Reisenbauer}, \citenamefont {Grass},
  \citenamefont {Kiesel}, \citenamefont {Vuleti{\' c}},\ and\ \citenamefont
  {Aspelmeyer}}]{delic_cavity_2018}%
  \BibitemOpen
  \bibfield  {author} {\bibinfo {author} {\bibfnamefont {Uro{\v s}}\
  \bibnamefont {Deli{\' c}}}, \bibinfo {author} {\bibfnamefont {Manuel}\
  \bibnamefont {Reisenbauer}}, \bibinfo {author} {\bibfnamefont {David}\
  \bibnamefont {Grass}}, \bibinfo {author} {\bibfnamefont {Nikolai}\
  \bibnamefont {Kiesel}}, \bibinfo {author} {\bibfnamefont {Vladan}\
  \bibnamefont {Vuleti{\' c}}}, \ and\ \bibinfo {author} {\bibfnamefont
  {Markus}\ \bibnamefont {Aspelmeyer}},\ }\bibfield  {title} {\enquote
  {\bibinfo {title} {Cavity cooling of a levitated nanosphere by coherent
  scattering},}\ }\href@noop {} {\bibfield  {journal} {\bibinfo  {journal}
  {arXiv:1812.09358 [physics, physics:quant-ph]}\ } (\bibinfo {year} {2018})},\
  \bibinfo {note} {arXiv: 1812.09358}\BibitemShut {NoStop}%
\bibitem [{\citenamefont {Clerk}\ \emph {et~al.}(2008)\citenamefont {Clerk},
  \citenamefont {Marquardt},\ and\ \citenamefont
  {Jacobs}}]{clerk_back-action_2008}%
  \BibitemOpen
  \bibfield  {author} {\bibinfo {author} {\bibfnamefont {A.~A.}\ \bibnamefont
  {Clerk}}, \bibinfo {author} {\bibfnamefont {F.}~\bibnamefont {Marquardt}}, \
  and\ \bibinfo {author} {\bibfnamefont {K.}~\bibnamefont {Jacobs}},\
  }\bibfield  {title} {\enquote {\bibinfo {title} {Back-action evasion and
  squeezing of a mechanical resonator using a cavity detector},}\ }\href
  {\doibase 10.1088/1367-2630/10/9/095010} {\bibfield  {journal} {\bibinfo
  {journal} {New Journal of Physics}\ }\textbf {\bibinfo {volume} {10}},\
  \bibinfo {pages} {095010} (\bibinfo {year} {2008})}\BibitemShut {NoStop}%
\bibitem [{\citenamefont {Suh}\ \emph {et~al.}(2014)\citenamefont {Suh},
  \citenamefont {Weinstein}, \citenamefont {Lei}, \citenamefont {Wollman},
  \citenamefont {Steinke}, \citenamefont {Meystre}, \citenamefont {Clerk},\
  and\ \citenamefont {Schwab}}]{suh_mechanically_2014}%
  \BibitemOpen
  \bibfield  {author} {\bibinfo {author} {\bibfnamefont {J.}~\bibnamefont
  {Suh}}, \bibinfo {author} {\bibfnamefont {A.~J.}\ \bibnamefont {Weinstein}},
  \bibinfo {author} {\bibfnamefont {C.~U.}\ \bibnamefont {Lei}}, \bibinfo
  {author} {\bibfnamefont {E.~E.}\ \bibnamefont {Wollman}}, \bibinfo {author}
  {\bibfnamefont {S.~K.}\ \bibnamefont {Steinke}}, \bibinfo {author}
  {\bibfnamefont {P.}~\bibnamefont {Meystre}}, \bibinfo {author} {\bibfnamefont
  {A.~A.}\ \bibnamefont {Clerk}}, \ and\ \bibinfo {author} {\bibfnamefont
  {K.~C.}\ \bibnamefont {Schwab}},\ }\bibfield  {title} {\enquote {\bibinfo
  {title} {Mechanically detecting and avoiding the quantum fluctuations of a
  microwave field},}\ }\href {\doibase 10.1126/science.1253258} {\bibfield
  {journal} {\bibinfo  {journal} {Science}\ }\textbf {\bibinfo {volume}
  {344}},\ \bibinfo {pages} {1262--1265} (\bibinfo {year} {2014})}\BibitemShut
  {NoStop}%
\bibitem [{\citenamefont {Shomroni}\ \emph {et~al.}(2018)\citenamefont
  {Shomroni}, \citenamefont {Qiu}, \citenamefont {Malz}, \citenamefont
  {Nunnenkamp},\ and\ \citenamefont {Kippenberg}}]{shomroni_optical_2018}%
  \BibitemOpen
  \bibfield  {author} {\bibinfo {author} {\bibfnamefont {Itay}\ \bibnamefont
  {Shomroni}}, \bibinfo {author} {\bibfnamefont {Liu}\ \bibnamefont {Qiu}},
  \bibinfo {author} {\bibfnamefont {Daniel}\ \bibnamefont {Malz}}, \bibinfo
  {author} {\bibfnamefont {Andreas}\ \bibnamefont {Nunnenkamp}}, \ and\
  \bibinfo {author} {\bibfnamefont {Tobias~J.}\ \bibnamefont {Kippenberg}},\
  }\bibfield  {title} {\enquote {\bibinfo {title} {Optical
  {Backaction}-{Evading} {Measurement} of a {Mechanical} {Oscillator}},}\
  }\href@noop {} {\bibfield  {journal} {\bibinfo  {journal} {arXiv:1809.01007
  [quant-ph]}\ } (\bibinfo {year} {2018})},\ \bibinfo {note} {arXiv:
  1809.01007}\BibitemShut {NoStop}%
\bibitem [{\citenamefont {Kerdoncuff}\ \emph {et~al.}(2015)\citenamefont
  {Kerdoncuff}, \citenamefont {Hoff}, \citenamefont {Harris}, \citenamefont
  {Bowen},\ and\ \citenamefont
  {Andersen}}]{kerdoncuff_squeezing-enhanced_2015}%
  \BibitemOpen
  \bibfield  {author} {\bibinfo {author} {\bibfnamefont {Hugo}\ \bibnamefont
  {Kerdoncuff}}, \bibinfo {author} {\bibfnamefont {Ulrich~B.}\ \bibnamefont
  {Hoff}}, \bibinfo {author} {\bibfnamefont {Glen~I.}\ \bibnamefont {Harris}},
  \bibinfo {author} {\bibfnamefont {Warwick~P.}\ \bibnamefont {Bowen}}, \ and\
  \bibinfo {author} {\bibfnamefont {Ulrik~L.}\ \bibnamefont {Andersen}},\
  }\bibfield  {title} {\enquote {\bibinfo {title} {Squeezing-enhanced
  measurement sensitivity in a cavity optomechanical system},}\ }\href
  {\doibase 10.1002/andp.201400171} {\bibfield  {journal} {\bibinfo  {journal}
  {Annalen der Physik}\ }\textbf {\bibinfo {volume} {527}},\ \bibinfo {pages}
  {107--114} (\bibinfo {year} {2015})},\ \bibinfo {note} {arXiv:
  1611.09772}\BibitemShut {NoStop}%
\bibitem [{\citenamefont {Mason}\ \emph {et~al.}(2018)\citenamefont {Mason},
  \citenamefont {Chen}, \citenamefont {Rossi}, \citenamefont {Tsaturyan},\ and\
  \citenamefont {Schliesser}}]{mason_continuous_2018}%
  \BibitemOpen
  \bibfield  {author} {\bibinfo {author} {\bibfnamefont {David}\ \bibnamefont
  {Mason}}, \bibinfo {author} {\bibfnamefont {Junxin}\ \bibnamefont {Chen}},
  \bibinfo {author} {\bibfnamefont {Massimiliano}\ \bibnamefont {Rossi}},
  \bibinfo {author} {\bibfnamefont {Yeghishe}\ \bibnamefont {Tsaturyan}}, \
  and\ \bibinfo {author} {\bibfnamefont {Albert}\ \bibnamefont {Schliesser}},\
  }\bibfield  {title} {\enquote {\bibinfo {title} {Continuous {Force} and
  {Displacement} {Measurement} {Below} the {Standard} {Quantum} {Limit}},}\
  }\href@noop {} {\bibfield  {journal} {\bibinfo  {journal} {arXiv:1809.10629
  [quant-ph]}\ } (\bibinfo {year} {2018})},\ \bibinfo {note} {arXiv:
  1809.10629}\BibitemShut {NoStop}%
\bibitem [{\citenamefont {Filip}\ and\ \citenamefont
  {Rakhubovsky}(2015)}]{filip_transfer_2015}%
  \BibitemOpen
  \bibfield  {author} {\bibinfo {author} {\bibfnamefont {Radim}\ \bibnamefont
  {Filip}}\ and\ \bibinfo {author} {\bibfnamefont {Andrey~A.}\ \bibnamefont
  {Rakhubovsky}},\ }\bibfield  {title} {\enquote {\bibinfo {title} {Transfer of
  non-{Gaussian} quantum states of mechanical oscillator to light},}\ }\href
  {\doibase 10.1103/PhysRevA.92.053804} {\bibfield  {journal} {\bibinfo
  {journal} {Physical Review A}\ }\textbf {\bibinfo {volume} {92}},\ \bibinfo
  {pages} {053804} (\bibinfo {year} {2015})}\BibitemShut {NoStop}%
\bibitem [{\citenamefont {Gieseler}\ and\ \citenamefont
  {Millen}(2018)}]{gieseler_levitated_2018}%
  \BibitemOpen
  \bibfield  {author} {\bibinfo {author} {\bibfnamefont {Jan}\ \bibnamefont
  {Gieseler}}\ and\ \bibinfo {author} {\bibfnamefont {James}\ \bibnamefont
  {Millen}},\ }\bibfield  {title} {\enquote {\bibinfo {title} {Levitated
  {Nanoparticles} for {Microscopic} {Thermodynamics} - a {Review}},}\ }\href
  {\doibase 10.3390/e20050326} {\bibfield  {journal} {\bibinfo  {journal}
  {Entropy}\ }\textbf {\bibinfo {volume} {20}},\ \bibinfo {pages} {326}
  (\bibinfo {year} {2018})},\ \bibinfo {note} {arXiv: 1805.02927}\BibitemShut
  {NoStop}%
\bibitem [{\citenamefont {Eisert}\ \emph {et~al.}(2002)\citenamefont {Eisert},
  \citenamefont {Scheel},\ and\ \citenamefont
  {Plenio}}]{eisert_distilling_2002}%
  \BibitemOpen
  \bibfield  {author} {\bibinfo {author} {\bibfnamefont {J.}~\bibnamefont
  {Eisert}}, \bibinfo {author} {\bibfnamefont {S.}~\bibnamefont {Scheel}}, \
  and\ \bibinfo {author} {\bibfnamefont {M.~B.}\ \bibnamefont {Plenio}},\
  }\bibfield  {title} {\enquote {\bibinfo {title} {Distilling {Gaussian}
  {States} with {Gaussian} {Operations} is {Impossible}},}\ }\href {\doibase
  10.1103/PhysRevLett.89.137903} {\bibfield  {journal} {\bibinfo  {journal}
  {Physical Review Letters}\ }\textbf {\bibinfo {volume} {89}},\ \bibinfo
  {pages} {137903} (\bibinfo {year} {2002})}\BibitemShut {NoStop}%
\bibitem [{\citenamefont {Fiur{\' a}{\v s}ek}(2002)}]{fiurasek_gaussian_2002}%
  \BibitemOpen
  \bibfield  {author} {\bibinfo {author} {\bibfnamefont {Jarom\'{\i}r}\
  \bibnamefont {Fiur{\' a}{\v s}ek}},\ }\bibfield  {title} {\enquote {\bibinfo
  {title} {Gaussian {Transformations} and {Distillation} of {Entangled}
  {Gaussian} {States}},}\ }\href {\doibase 10.1103/PhysRevLett.89.137904}
  {\bibfield  {journal} {\bibinfo  {journal} {Physical Review Letters}\
  }\textbf {\bibinfo {volume} {89}},\ \bibinfo {pages} {137904} (\bibinfo
  {year} {2002})}\BibitemShut {NoStop}%
\bibitem [{\citenamefont {Serafini}(2017)}]{serafini_quantum_2017}%
  \BibitemOpen
  \bibfield  {author} {\bibinfo {author} {\bibfnamefont {Alessio}\ \bibnamefont
  {Serafini}},\ }\href@noop {} {\emph {\bibinfo {title} {Quantum {Continuous}
  {Variables}: A {Primer} of {Theoretical} {Methods}}}}\ (\bibinfo  {publisher}
  {CRC Press},\ \bibinfo {year} {2017})\ \bibinfo {note} {google-Books-ID:
  bMItDwAAQBAJ}\BibitemShut {NoStop}%
\end{thebibliography}%
\end{document}